\documentclass[12pt,preprint]{aastex}
\shorttitle{Stellar populations in spiral nuclei}
\shortauthors{Walcher et al.}

\begin{document} 

\title{Stellar populations in the nuclei of late-type spiral galaxies \footnote{Based on observations collected at the European Southern Observatory, Chile, proposal No. 68.B-0076} }

\author{C.~J. Walcher\altaffilmark{1}, 
T. B\"oker\altaffilmark{2},
S. Charlot\altaffilmark{3,4},
L.~C. Ho\altaffilmark{5},
H.-W. Rix\altaffilmark{1},
J. Rossa\altaffilmark{6}, 
J.~C. Shields\altaffilmark{7},
R.~P. van der Marel\altaffilmark{6}}

\altaffiltext{1}{Max Planck Institut f\"ur Astronomie, K\"onigstuhl 17, D-69117 Heidelberg, Germany}
\altaffiltext{2}{ESA/ESTEC, Keplerlaan 1, 2200 AG Noordwijk, Netherlands}
\altaffiltext{3}{Max Planck Institut f\"ur Astrophysik, Karl-Schwarzschild-Str. 1, D-85741 Garching, Germany}
\altaffiltext{4}{Institut d'Astrophysique de Paris, 98bis, bd Arago, F-75014 Paris, France}
\altaffiltext{5}{Observatories of the Carnegie Institution of Washington, 813 Santa Barbara Street, Pasadena, CA 91101-1292, USA}
\altaffiltext{6}{Space Telescope Science Institute, 3700 San Martin Drive, Baltimore, MD 21218, USA}
\altaffiltext{7}{Department of Physics and Astronomy, Ohio University, Athens, OH45701-2979, USA}

\begin{abstract}

As part of an ongoing effort to study the stellar nuclei of very 
late-type, bulge-less spirals, we present results from 
a high-resolution spectroscopic survey of nine such 
nuclear star clusters, undertaken with VLT/UVES. 
We fit the spectra with population synthesis models and 
measure Lick-type indices to determine mean 
luminosity-weighted ages, which range from $4.1\times 10^7$ 
to $1.1\times 10^{10}$ years and are insensitive to assumed 
metallicity or internal extinction. The average metallicity of 
nuclear clusters in late-type spirals is slightly sub-solar 
($\langle Z \rangle = 0.015$) but shows significant scatter.
Most of the clusters have moderate extinctions of 0.1 to 0.3 mags 
in the $I$-band. The fits also show that the nuclear cluster spectra 
are best described by a mix of several generations of stars. This is 
supported by the fact that only models with composite stellar populations 
yield mass-to-light ratios that match those obtained from dynamical
measurements. For our nine sample clusters, 
the last star formation episode was on average 34 Myr ago, while 
all clusters experienced some star formation in the last 100 Myr. 
We thus conclude that the nuclear clusters undergo repeated 
episodes of star formation. The robustness of our results with respect 
to possible contamination from the underlying galaxy disk is 
demonstrated by comparison to a similar analysis using 
smaller-aperture spectra obtained with HST/STIS.
Combining these results with those from Walcher et al. 
(2005), we have thus shown that the stellar nuclei of these 
bulge-less galaxies are massive and dense star clusters that form 
stars recurrently until the present day. This set of properties 
is unique among the various classes of star clusters. 
It is almost inevitable to associate these unique properties 
with the location of the cluster in its host galaxy. It 
remains a challenging question to elucidate exactly how 
very late-type spirals manage to create nuclei with such 
extreme characteristics.

\end{abstract}

\keywords{galaxies: star clusters; galaxies: nuclei; galaxies: structure; 
galaxies: spiral}

%%%%%%%%%%%%%%%%%%%%%%%%%%%%%%%%%%%%%%%%%%%%%%%%%%%%%%%%%%%%%
\section{Introduction}
\label{sect:intro}

Galaxy centers have attracted special interest from astronomers 
for a long time, as they are the places of a number of distinctive 
phenomena, such as active galactic nuclei, central starbursts 
and extremely high stellar densities. The last decade has also shown 
that the evolution of galaxies is closely linked to the evolution 
of their nuclei, as evidenced by a number of global to nucleus 
relations (Ferrarese \& Merrit 2000; Gebhardt et al. 2000; 
Graham et al. 2001; Ferrarese 2002; H\"aring \& Rix 2004). 
In view of this general paradigm we here present our ongoing 
efforts to study the nuclear star clusters in late-type 
spirals as a contribution to a full census of galaxy nuclei 
over all Hubble types.

It has been thought for 20 years that ``the centers of some, 
perhaps most, galaxies contain a nuclear stellar system which 
is dynamically distinct from the surrounding bulge or disk 
components'' (O'Connell 1983). Indeed, in our close vicinity, 
nuclear star clusters (NCs) are found in the Galactic center 
(Krabbe et al. 1995; Genzel \& Eckart 1998), in M31 
(Johnson 1961; Davidge 1997; Lauer et al. 1998), and in M33 
(Gallagher, Goad \& Mould 1982). We here reserve the word NC for 
the one central star cluster of a galaxy, which is in the dynamic 
center of the galaxy, i.e. at the bottom of its potential well. 
On a larger scale, Phillips et al. (1996) and Matthews \& Gallagher 
(1997) surveyed the nuclei of nearby spirals and found NCs in 6 out of 10 
and in 10 out of 49 late-type spirals, respectively 
(see also Matthews et al. 1999). 
Subsequently, Carollo et al. (1997) and Carollo, Stiavelli \& Mack 
(1998) found nuclear point-like sources to be present in many 
of the nearby spiral galaxies in their sample. Further down the Hubble 
sequence exist such examples as the bright, central cluster in 
NGC 1705 (Ho \& Filippenko, 1996), a dwarf irregular galaxy. 
Also active galaxies 
are observed to host NCs in some cases (e.g., Thatte et al. 1997; 
Gallimore \& Matthews 2003; Schinnerer et al. 2001).
In the specific case of late-type spirals, 
B\"oker et al (2002, hereafter B02) found that 
NCs are present in $\approx$75\% of the galaxies. 

Although these clusters are almost ubiquitous at least in spiral 
galaxies, their properties have not been extensively studied. 
Indeed, for all galaxies with more than a tiny bulge component 
difficulties such as extinction and contamination from bulge 
light make it difficult to study the NCs 
as distinct entities. For the latest Hubble types, 
several case studies in the literature, however, have 
revealed compact, massive, young objects. Typical sizes 
are less than 5 pc (see e.g. B\"oker et al. 2004, hereafter B04). 
M33 is the nearest Sc galaxy hosting a NC and its nucleus has been extensively 
studied in the past decade (recently e.g. in Davidge 2000; Long, Charles 
\& Dubus 2002; Stephens \& Frogel 2002). Despite some differences in the 
details, all studies of M33 agree that there is some population younger than 
0.5 Gyr in the central parsec and that star formation has varied significantly 
over the past several Gyrs. The mass of the central cluster in M33 was 
estimated from a detailed population analysis in Gordon et al. (1999) to 
be $5\times 10^5$ M$_{\sun}$, consistent with the upper limit derived 
from the velocity dispersion by Kormendy \& McClure (1993) of $2\times 
10^6$ M$_{\sun}$. For the NC in NGC 1705, Ho \& Filippenko (1996) measured the 
velocity dispersion and used the virial theorem to derive 
a mass of $8.2 \times 10^4$  M$_{\sun}$. 
From H$\alpha$ rotation curves of late-type spirals, Matthews \& Gallagher 
(2002) find that the velocity offsets at the position of five semi-stellar
nuclei --- certainly to be identified with NCs --- 
are consistent with masses of $\approx 10^6$ -- $10^7$ M$_{\sun}$. 
They also point out that the location of the cluster and the 
dynamical center of the galaxy do not always coincide. The only direct mass 
determination from a measurement of the stellar velocity 
dispersion and detailed dynamical modeling was done for the NC in IC342
by B\"oker, van der Marel \& Vacca (1999). They determined a mass of 
$6\times 10^6$ M$_{\sun}$, with a $K$-band mass-to-light ratio of 0.05 
in solar units. The derived mean age is $\le 10^8$ 
years. B\"oker et al. (2001) also studied the NC in NGC~4449 using population 
synthesis models. They estimate an age of $\approx 10^7$ years 
in agreement with Gelatt, Hunter \& Gallagher (2001). They infer a lower 
limit for the mass, which is $4\times 10^5$ M$_{\sun}$. 

Most recently, Walcher et al. (2005, 
hereafter Paper I) have directly measured the 
dynamical masses of nine NCs from the 
sample of B02. The main results were that although NCs 
follow typical scaling relations of other compact star 
clusters, e.g. globular clusters, they stand out 
as being among the most massive stellar clusters known 
with a typical mass around $5\times 10^6$ M$_{\sun}$. 
Their small sizes however place them far away from low-luminosity 
``spheroids'' (e.g. bulges or dE galaxies). Though apparently inconspicuous, 
the centers of the majority of late-type spirals must 
have a way to form exactly one massive, dense cluster. 
It is not clear why that should be so, as rotation curve measurements 
of galaxies of similar type show that their 
gravitational potentials are shallow, i.e. the gravitational 
vector vanishes at the center (e.g. de Blok et al. 2001a, 2001b, 
Matthews \& Gallagher 2002, Marchesini et al. 2002). Why the 
center of these galaxies should be a special place is thus not
easily seen. Weidner, Kroupa \& Larsen (2004) have discussed the 
possible existence of an upper mass limit for bound stellar clusters
of around $5 \times 10^6$ M$_{\sun}$ in the sense that clusters 
above this limit 
are observed to have complex stellar populations that cannot be 
the result of a single star formation event. 
It is interesting to note in this context 
that all clusters that lie above or near this possible upper mass limit 
either presently 
lie at the center of a galaxy (dE nuclei, NC), or are 
thought to have formed in such a galaxy center, the rest of 
the surrounding galaxy having been stripped away by tidal forces. 
This scenario has been invoked for $\omega$Cen (e.g., 
Bekki \& Freeman 2003), G1 (e.g. Bekki \& Chiba 2004), 
and Ultra Compact Dwarfs (Drinkwater et al. 2003). 

Clearly, in order to understand the formation processes for such NCs, 
a second clue next to size and dynamical mass are the age and 
metallicity of the cluster. 
In this paper we thus complement the dynamical study presented in 
Paper I by a study of the stellar populations 
in the nuclei of nine extreme late-type spiral galaxies.

%%%%%%%%%%%%%%%%%%%%%%%%%%%%%%%%%%%%%%%%%%%%%%%%%%%%%%%%%%%%%
\section{UVES: Sample Selection and Data Reduction}
\label{s:UVES}

The data, a set of high spectral resolution integrated cluster 
spectra, used in this study have been presented in Paper I. 
We repeat only some central points 
with special emphasis on the blue parts of the spectra. 

\subsection{Sample Selection}
All objects were drawn from the sample of B02. Galaxies 
in this survey were selected to have Hubble type between Scd and Sm, 
line-of-sight velocity v$_{\mbox{hel}} <$ 2000 km~s$^{-1}$ and to be 
close to face-on. NCs were found in the photometric centers of 75\% 
of the galaxies as luminous, barely resolved sources. We emphasize that 
there is always only one nuclear cluster per galaxy, which is typically 
two magnitudes brighter than all other clusters in the galaxy (B02). 
Spectroscopic data for 9 clusters were taken with the Ultraviolet and Visual 
Echelle Spectrograph (UVES) at the VLT. The objects were 
selected from the full catalogue to be accessible during the time 
of observations and to be bright 
enough to be observed in less than three hours, thus maximizing the 
number of observable objects. We thus sample the brighter 2/3 of the 
luminosity range covered by the clusters. Whether this bias in absolute 
magnitude introduces others, e.g. towards younger or more massive 
clusters, is at present unclear (but see Rossa et al. 2006).

\subsection{Observations}
The VLT spectra were taken with the UVES 
Spectrograph attached to Kueyen (UT2) in December 2001. 
All nights were clear with a seeing around $1''$. UVES provides us 
with two wavelength regions, namely 3570-4830 {\AA}  in the blue arm, 
while the red arm is split on two CCDs, the lower one covering 
6120-7980 {\AA} and the upper one covering 8070-9920 {\AA}. The slit was 
centered on the position of the NC as derived from 
the HST images. The length of the slit was $10''$, the width $1''$. 
We thus reach a resolution of R$\approx$32000 in the blue arm. 
The slit was always oriented perpendicular to the horizon to minimize 
effects of atmospheric refraction.

Basic properties of the observed NCs are listed 
in Table \ref{t:UVES}. For all objects, HST $I$-band images can be 
found in B02.

\subsection{Reduction}

The data were reduced with the UVES reduction pipeline version 1.2.0 
provided by ESO (Ballester et al., 2001). For a detailed 
description of our reduction steps, see Paper I.

Effective radii for the clusters were derived from HST photometry in 
B04. They are $\lesssim 0\farcs2$ for all observed 
clusters. Because the seeing is considerably larger than the effective 
radii of the clusters, we measure integrated properties of the whole 
cluster. However, because of the size of the slit there is also 
significant contamination from non-cluster light stemming from the 
galaxy disk in most of our spectra. The percentage of contamination 
from non-cluster 
light in the spectra was derived in Paper I and is quoted in 
Table \ref{t:UVES} as NCL (non-cluster light). NCL depends on the 
size of the extraction window in the sense that a larger window 
will lead to a larger NCL. However S/N is critical in the following 
analysis and was even more so in Paper I. Extraction windows were 
therefore optimized to yield as large a S/N as possible, rather than 
the smallest possible NCL, deferring 
the treatment of the NCL effects to the discussion.

Before computing the NCL, we subtracted from the nucleus spectrum the
background measured from adjacent regions of the spectrograph slit,
which samples the night sky as well as starlight in the surrounding
disk. As explained in Paper I, the host galaxy surface brightness
distribution is not expected to be flat across the central region
where the NC is located. Not all of the disk light is therefore 
subtracted. The NCL values represent our estimate of
the resulting residuals.
We decided to proceed without any attempt to further correct 
for the disk contamination for the two following reasons: 
1) The inward extrapolation of the disk luminosity profile to the 
center of the cluster is uncertain (see B02). Accurate subtraction 
of the disk spectrum is therefore not possible.
2) The disk spectrum has low S/N. If we subtracted a 
considerable amount of such a disk spectrum, the resulting 
spectrum would have S/N similar to the disk spectrum - 
which would make all the following analysis impossible. 

Two different spectrophotometric standard stars were observed 
on two of the three nights. The two response 
curves we obtain on two different nights and from two different 
stars agree to a level of approximately 2\%. We corrected the final 
reduced object spectra for atmospheric extinction with the extinction curve 
for La Silla found at the web page of ESO. 
We made no attempt to check if the absolute flux calibration of the 
spectra is correct, as we are only interested in the relative flux 
calibration for the purposes of this paper.

\begin{figure}[tbp]
\begin{center}
\includegraphics[width=1\hsize]{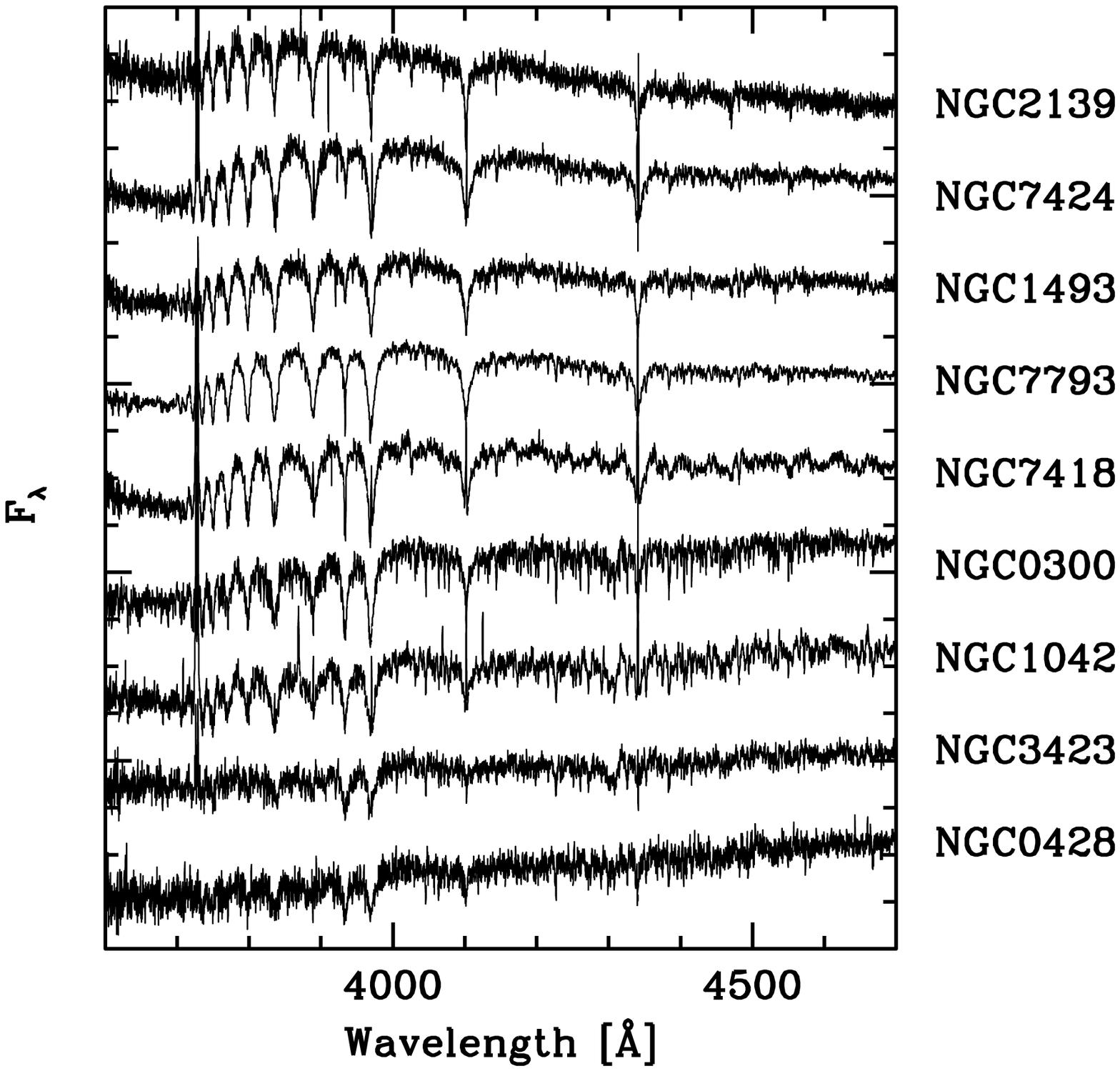}
\end{center}
\caption[all_blue]{Reduced, response-corrected blue spectra 
of the nuclear regions ($\sim 1''$) of the nine galaxies in our sample. 
Mean stellar ages increase from top to bottom. The pseudo-flux 
scale and offset of each spectrum have been adjusted arbitrarily.}
\label{f:all_blue}
\end{figure}

Figure \ref{f:all_blue} shows the reduced, response and atmospheric 
extinction corrected, 
sky-subtracted spectra of all nine nuclear regions on the blue 
CCD chip. For presentation purposes the pseudo-flux scale and 
offset of each spectrum have been adjusted arbitrarily. 
The spectra are ordered approximately in the age sequence 
that will be derived later, with the youngest on top.

%%%%%%%%%%%%%%%%%%%%%%%%%%%%%%%%%%%%%%%%%%%%%%%%%%%%%%%%%%%%%
\section{Determining the Star Formation Histories of Nuclear Clusters}
\label{s:age}

In the following we will apply three different measures, namely 
spectral index analysis, fitting single-age stellar population 
spectra, and fitting composite population spectra, to quantify 
the stellar population ages of the nuclei. This procedure will 
not only allow us to cross-check the methods against each other, 
but it will also yield the maximum amount of information 
available from the spectra and the current population synthesis 
models. 

We start by using spectral indices (see Section \ref{s:lick}) 
to derive the ages of our nuclear spectra. This is a widely used 
method, designed to compress the information available in the 
spectrum into a small set of numbers. In Sections 
\ref{s:SSP} and \ref{s:mixpop} we then also fit a larger wavelength 
range with model spectra to investigate whether the observed spectra 
are well described by single-age or multicomponent models. 
Yet another method which is beyond 
the scope of the present paper, but that can be found in the 
literature, uses integrated spectra of actual clusters with well 
determined ages and metallicities as templates (see e.g. Bica et al. 1998).

Population synthesis models exploit the fact 
that stellar populations with any star formation history can be 
expanded in a series of instantaneous starbursts, conventionally 
named simple stellar populations (SSP, see e.g. Bruzual \& 
Charlot 2003 and references therein). Good examples of well studied 
populations of coeval star clusters are globular clusters 
or very young star clusters, as formed during starburst events. 
The Spectral Energy Distribution (SED) of any unreddened stellar 
population at time $t$ can then be written as 

\begin{equation}
 F_{\lambda}(t) = \int_{0}^{t} \Psi (t-t') S_{\lambda}[t',\zeta (t-t')] dt'.
\label{eq:sedmix}
\end{equation}

Here, $\Psi (t-t')$ is the star formation rate over time, $\zeta (t-t')$ 
is the metal-enrichment law and $S_{\lambda}[t,\zeta (t-t')]$ is the 
power radiated per unit wavelength per unit initial mass by an SSP 
of age $t'$ and metallicity  $\zeta (t-t')$. 
It follows from this equation that by choosing suitably spaced 
SSPs covering a range of ages and metallicities, the problem 
of solving for the star formation history of a stellar system 
is equivalent to defining and minimizing the merit function 
\begin{equation}
 \chi^2 = \sum_{i=0}^{n} \left[ \frac{F_i - \sum_{k=1}^{M} a_k S_{i}[t_k,Z]}{\sigma_i} \right]^2, 
\label{eq:merit}
\end{equation}
over all non-negative $a_k$. Here $F_i$ is the observed spectrum in 
each of $n$ wavelength bins $i$, $\sigma_i$ is the standard deviation 
and $a_k$ are weights attributed to each of $M$ SSP 
models $S_{i}[t_k,Z]$ of age $t_k$ and metallicity $Z$. This merit 
function is linear in the $a_k$. In theory, the full SED thus 
contains all available constraints on all the different 
parameters of the system. Below we will introduce the same code 
used in Paper I for the kinematic analysis of the nuclear spectra, now used 
as a tool to derive the best fitting linear combination of a suitable 
set of SSP template spectra. This approach uses the SED over much 
of the wavelength range, in contrast to the spectral indices from the 
Lick/IDS system, which are designed to focus on specific parts of the 
spectra and thus compress the available information. As will be seen, 
the spectral fitting method can be used to 
derive ages, metallicities, extinctions and the age of the last 
star formation burst for the spectra in our sample.

We use the population synthesis model presented by Bruzual \& Charlot 
(2003, hereafter BC03). This model has a number of desirable features 
for the following analysis, which will be described 
briefly. The model covers a wide wavelength range from 3200 
to 9500 {\AA} for a wide range of metallicities (from $Z$ = 0.0004 
to 0.05, where 0.02 is solar). In the following only metallicities 
from 0.004 to 0.05 are used, as lower metallicities are not well 
sampled by the empirical stellar library used for predicting 
SEDs (STELIB, {Le Borgne et al.}, 2003). The models assume solar
abundance ratios. These ratios, e.g. $\alpha$-elements to Iron are, 
however, known to vary in external galaxies; this is 
a limitation to our analysis. As the observed spectra 
reach from 3600 to 9900 {\AA} (albeit with a gap from 4800 to 
6150 {\AA}) such a large wavelength coverage is desirable. The 
model SEDs cover an age range from $10^5$ to $2 \times 10^{10}$ 
years. Model uncertainties become large for ages younger than 
$10^6$ years and we will therefore restrain the used age range 
to SED ages larger than 1 Myr. If formation timecales for star 
clusters are of the order of $10^6$ years (as argued by Weidner 
et al. 2004), and therefore some of the stars are 
still in the contraction phase, the assumption that all stars have 
reached the zero-age main sequence might not be justified. 
Further, such very young stellar populations 
can be expected to be still embedded in the original 
gas cloud from which they formed and thus to be heavily reddened 
or even fully extinguished. The SEDs have a resolution of 
3 {\AA} FWHM across the whole wavelength range, which corresponds 
to a median $\lambda/\Delta \lambda = 2000$ . While this 
resolution is lower than that of our observations, the BC03 
models are the best available for matching the characteristics 
of our data. Only one model has become 
available recently that has a higher resolution of R $\approx$ 10000 
({Le Borgne et al.}, 2004). However 
the SEDs cover a somewhat smaller wavelength range (4000 to 6800 
{\AA}) and do not predict SEDs for ages younger than $10^7$ years.

The initial mass function (IMF) is a free parameter of the model and we 
follow BC03 in choosing a Chabrier (2003b) IMF. The spectral 
properties obtained using the above IMF are similar to those obtained 
using the Kroupa (2001) IMF. BC03 adopted the Chabrier 
IMF because it is physically motivated and provides a better fit 
to counts of low-mass stars and brown dwarfs in the Galactic disc 
(Chabrier 2001, 2002, 2003a).
The SED of each model SSP is normalized to an initial total 
mass in stars of 1 M$_{\odot}$. Thus a 10 Gyr old SSP will 
represent only about 0.5 M$_{\odot}$ in stars, with the rest 
of the initial mass lost through stellar winds and supernova 
explosions.

%%%%%%%%%%%%%%%%%%%%%%%%%%%%%%%%%%%%%%%%%%%%%%%%%%%%%%%%%%%%%
\subsection{Ages from Spectral Indices}
\label{s:lick}

To determine characteristic ages and metallicities of the observed 
spectra we first use Lick/IDS indices (Worthey et al.~1994,
Worthey \& Ottaviani 1997, Trager et al.~1998), originally designed 
to resolve the age-metallicity degeneracy for old stellar systems. 
The stellar population models predict the values of a set of absorption 
feature indices for each metallicity and single-burst age. For the 
population synthesis model of BC03 a set of 31 spectral indices is 
available in tabulated form. 
By comparing these modeled indices to those measured in 
observed spectra, one can estimate the age of the stellar cluster or 
galaxy under study.

\begin{figure}[t]
\begin{center}
\parbox{1\hsize}{
\includegraphics[angle=0,clip,width=0.48\hsize]{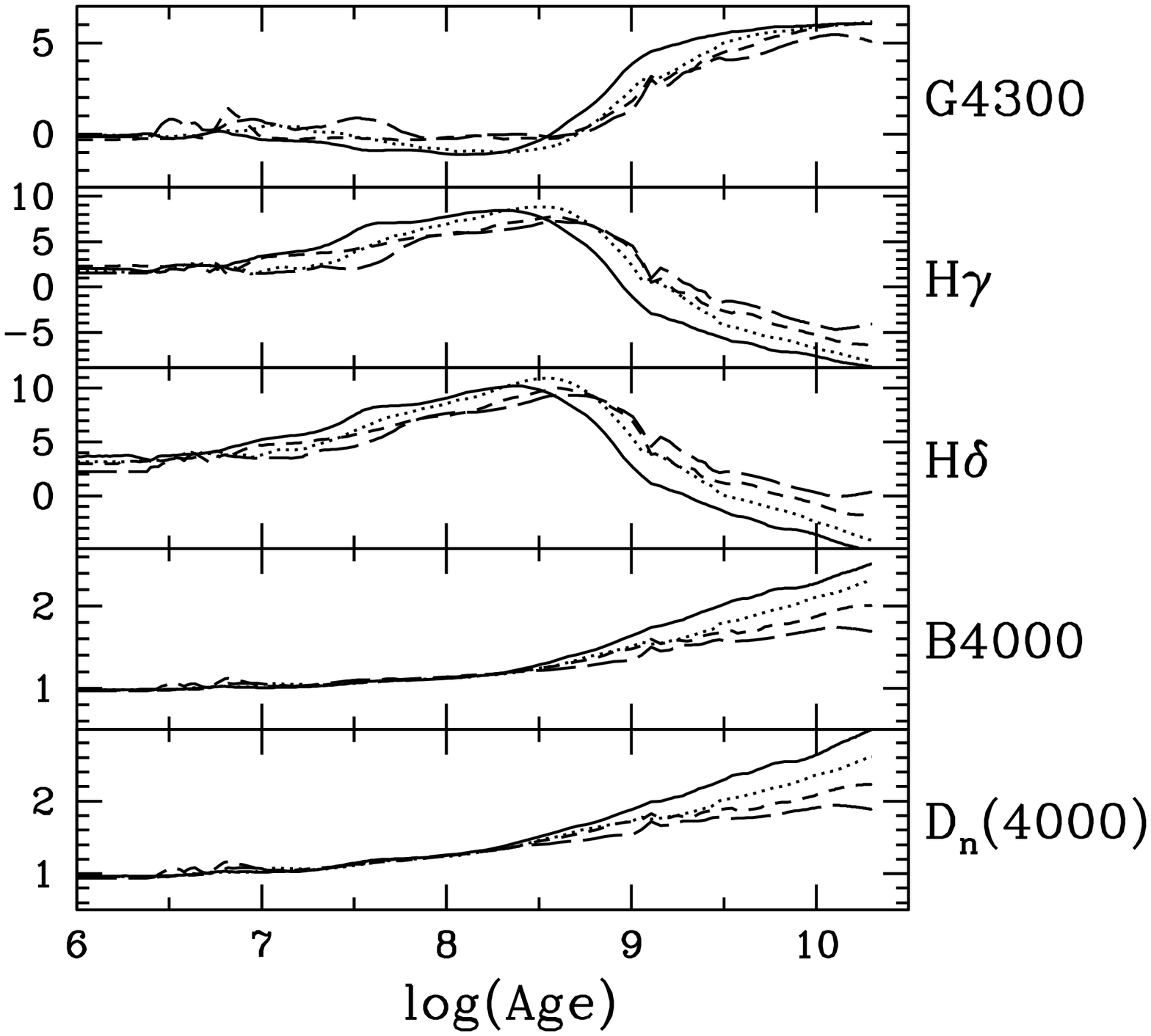}
\includegraphics[angle=0,clip,width=0.48\hsize]{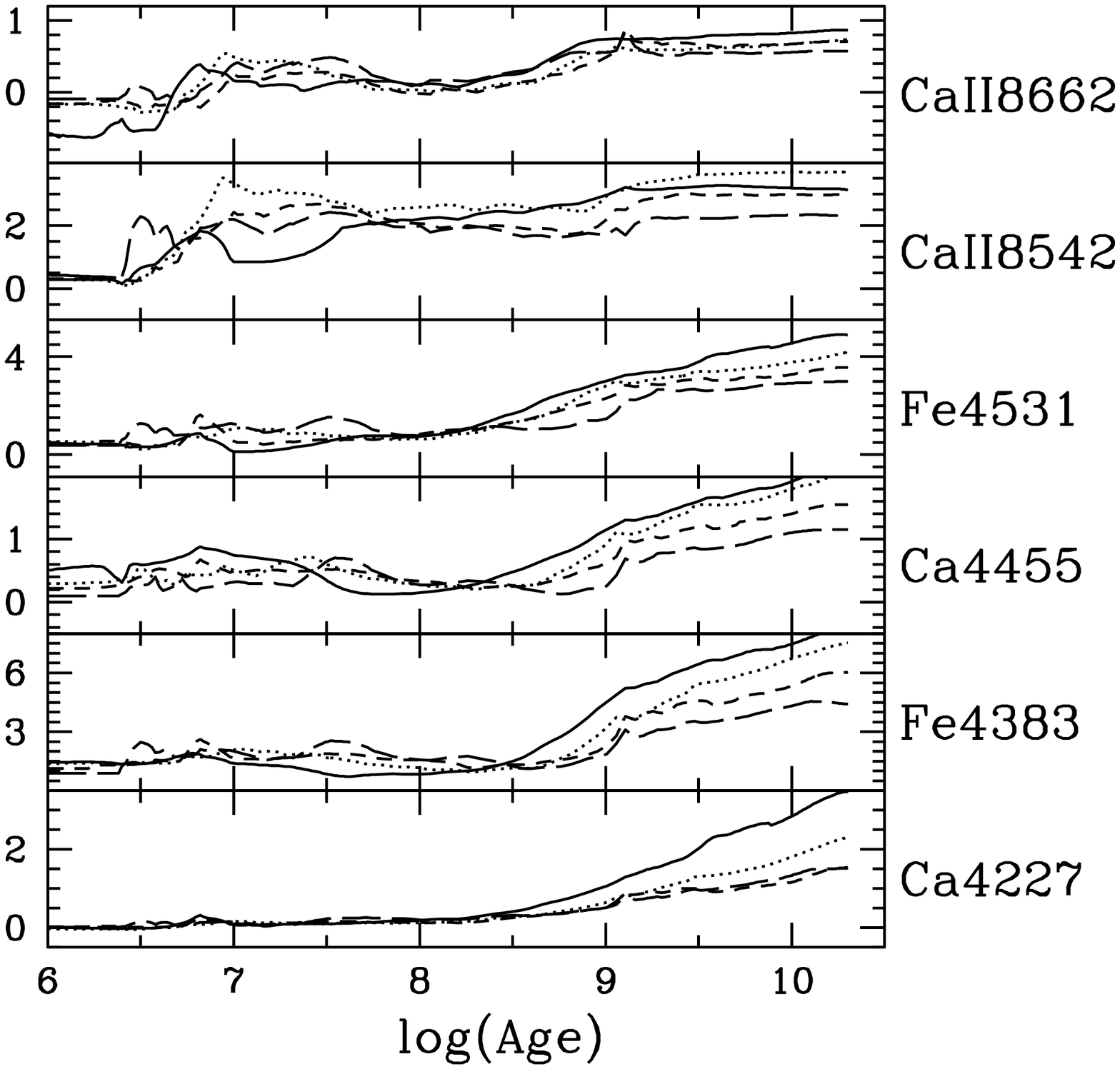}}
\end{center}
\caption[overview]{Dependence on age and metallicity of different spectral 
indices measurable in the UVES data as predicted from the 
BC03 models. Each line illustrates how the indices behave as the
model age is varied, and the different lines correspond to 
different metallicities (solid $Z$=0.05, dotted $Z$=0.02, short 
dashed $Z$=0.008, long dashed $Z$=0.004). Note that the left five 
indices are good age indicators (though the Balmer indices are double 
valued, so require a further constraint for age measurements). The six 
indices on the right are more sensitive to metallicity, again in 
particular at old ages.}
\label{f:indices}
\end{figure}

For the present work, we cannot use the full system of Lick 
indices for several reasons. First, many of the indices 
lie in spectral regions not covered by our data. Among them are 
most of the Fe indices (Fe5270, Fe5335, Fe5406, Fe5709 and Fe5782) 
and the strongest age indicator, namely the H$\beta$ index at 4861 {\AA}.
Second, emission lines, which are not accounted for in the models, 
are present in the spectra. They partially 
fill up the Balmer absorption troughs and thus bias age measurements 
to higher ages. Third, the index libraries have been designed mostly 
for populations older than 1 Gyr. Especially the metallicity of 
young populations is not well constrained by most of the indices. 
This statement is illustrated in Figure \ref{f:indices} that shows 
all the indices that we measure in our data as a function of 
metallicity and age. For young populations, the index values do not 
depend heavily either on age or on metallicity, making it difficult 
to use them, given the noise in the data. 
Further the spectra of very young/hot stars 
are dominated by electron scattering, and therefore all metal-absorption
lines are extremely weak in the wavelength range we have access to.
Although in our wavelength range, the CN 
indices will not be considered here as they fall into the family of 
$\alpha$-element-like indices (as found by Trager et al.~1998) and are 
thus not a good measure of metallicity as defined by the [Fe/H] ratio. 
We also do not measure the third Ca index at 8498 {\AA} because of a 
CCD defect in this region of the spectrum. 

Indices were measured following the procedures outlined 
in Worthey et al. (1994). References for the indices we measure 
in this work are given in BC03 as well as the bandpasses 
and pseudo-continuum bands. Note that the indices were directly 
measured in the observed and in the modeled spectra in the same way 
(in particular at the same spectral resolution), 
without the use of fitting functions. 

The measured indices for our nine objects are tabulated in Table 
\ref{t:measind}. The spectra were degraded to the resolution 
of the models by convolving them with a Gaussian of 3 {\AA} FWHM 
and then rebinning them to the {1 {\AA}} pixel size. We have verified 
that the rebinning process changes the index values only slightly, 
i.e. less than the statistical 1 $\sigma$ error. This is negligible 
for our interpretation. Three of the spectra (NGC~1042, NGC~7418, 
NGC~7424) have prominent Balmer emission lines, which are less broad 
than the absorption lines. The emission lines have been subtracted 
by joining the two points of deepest absorption with a straight 
line prior to the degradation of resolution. 
Our goal is to compare the age determinations from the Lick 
indices with the direct fitting method introduced below. Therefore, 
emission line removal based on the fit of the BC03 models, though 
possible, would have led to circular reasoning and was not applied.
To allow for an assessment of the magnitude of the effect of removing 
the emission lines, we also quote the index values derived from the
non-interpolated spectra in Table \ref{t:measind}. This 
non-sophisticated procedure of emission line removal probably 
underestimates the equivalent width of the absorption lines. 

To compute errors on the indices 
we created 300 representations of the spectra, assuming Gaussian 
errors on each pixel and using the measured noise vectors. The error
on the index then is the rms of the 300 measured index values.

\begin{figure}[tbp]
\begin{center}
\includegraphics[angle=0,clip,width=0.9\hsize]{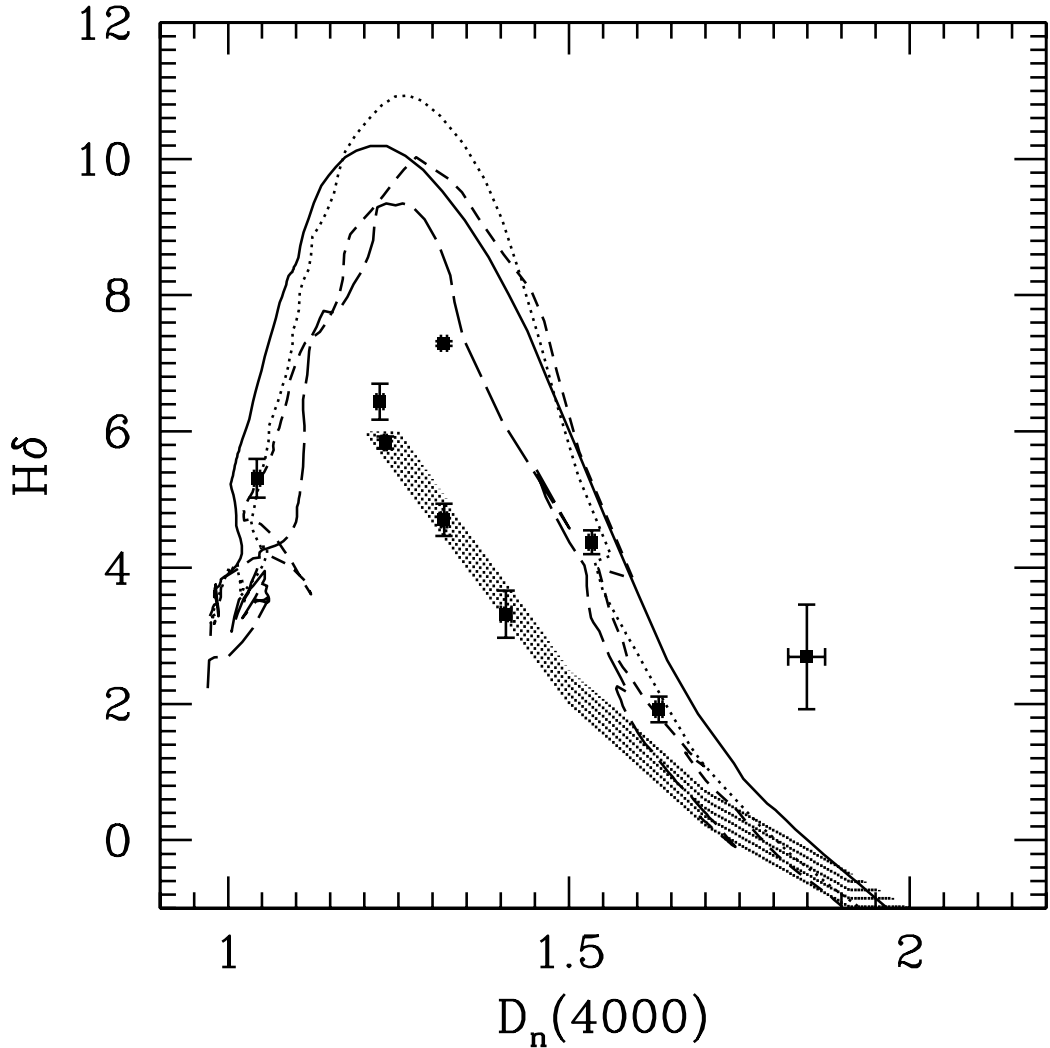}
\end{center}
\caption[measind]{Key spectral indices measured for the nine spectra and 
compared to the predictions by population synthesis models. 
We here plot two age sensitive indices, D$_n$(4000) and H$\delta_A$. 
The tracks of SSP models are indicated as different lines, corresponding 
to different metallicities (solid $Z$=0.05, dotted $Z$=0.02, short dashed 
$Z$=0.008, long dashed $Z$=0.004); models with continuous star formation 
are represented as a shaded area (from Kauffmann et al. 2003). 
Note that many NCs do not match the locus of the SSP models, but have 
indices indicative of continuous or repeated star formation.}
\label{f:metage}
\end{figure}

As Figure \ref{f:metage} shows, the measured index values in most cases 
do not match the tracks of the SSP models. 
The graph shows the predicted tracks of the indices 
from the BC03 models for two age indicator indices, namely the 
D$_n$(4000) index against the H$\delta_A$ index. The dependence of 
these indices on the star formation history has been studied 
quantitatively by Kauffmann et al. (2003) through large sets of 
simulations with different star formation histories. They find 
that continuous star formation histories occupy a narrow band 
(the shaded area in Figure \ref{f:metage}) in the D$_n$(4000)-H$\delta_A$ 
plane, remarkably close to the locus of many NCs. Stellar populations 
with stronger H$\delta_A$ absorption strength at a given value of 
D$_n$(4000) must have formed a significant fraction of their stars 
in a recent burst. Depending on burst-strength, they fill the 
space between continuous star formation and the single burst 
models (SSPs). Thus the line indices alone indicate that the spectra 
are not well represented by a single-age population. Note that dust can affect
both the D$_n$(4000) and H$\delta_A$ indices quite significantly such that
an intrinsically continuous star formation history could be moved towards the
"bursty" region of the grid for reasonable amounts of dust
(see MacArthur 2005, Fig. 13). However, due to the face-on orientation of 
the galaxies in our sample, intrinsic reddening should be small (cf. 
Section \ref{s:age_methods}). The one object 
that falls out of the region covered by SSPs and continuous star formation 
histories alike in Figure \ref{f:metage} is NGC 428 at a D$_n$(4000) value of 
1.85. This could possibly arise from the extinction effect described in 
MacArthur (2005), as we indeed find an unusually high extinction for 
this NC in Section \ref{s:mixpop}. 
Finally, although we quote both indices for the sake 
of completeness, the D$_n$(4000) index differs only slightly in definition 
from the B(4000) index, both measure the strength of the 4000 {\AA} break. 
Also, the two Balmer line indices H$\delta_A$ and H$\gamma$ are mostly 
redundant. 

Typically, a set of two indices, one age-dependent, one metallicity-dependent, 
can be used to break the 
age-metallicity degeneracy and derive an SSP age from an observed spectrum. 
However, as seen in the previous paragraph, 
in our case we need at least two age indices for the age determination 
alone, because our objects cover a larger age range than typical studies 
of early type galaxies. Even though NCs are most probably {\emph not} SSPs, 
the best fit value for the age $\tau$ and the metallicity of the spiral 
nuclei is a useful benchmark for comparison purposes. We thus also 
examined the behavior of the more metal-dependent indices to identify 
those most likely to provide robust results. 

Although some of the indices that are commonly seen to be metallicity 
indicators, e.g. Ca4227, Fe4383, and Fe4531, are observed to be within 
the range of the model predictions, this is not the case for all the indices. 
For example the measured Ca\textsc{II}8662 index never matches the models 
(this is probably due to difficulties with the subtraction of night 
sky emission lines in this part of the spectrum). 
This casts doubt on the other Ca-triplet index Ca\textsc{II}8542 as well. 
The index Ca4455, though commonly used as a metallicity indicator, also 
depends significantly on age for the large age range spanned by our objects. 
The age range that is formally consistent 
with the measured value of Ca4455 is incompatible with the age range derived 
from the other indices (for any metallicity). Thus, we discard Ca4455 from 
the further analysis (compare also Thomas et al.~2003). 

Additional uncertainties to our analysis arise due to the lack of coverage 
of non-solar abundances in the models. Although exact values depend on the 
specific index, an increase of the [$\alpha$/Fe] ratio by 0.3 dex may lead 
to changes in the predicted index strengths by up to 0.5 {\AA}. It is 
conceivable that the use of models with different [$\alpha$/Fe] ratios 
would lead to an improved agreement between SSP model spectra and the 
measured index values.

Disregarding the most problematic and some redundant indices, leaves 
the following sub-set: B(4000), H$\gamma$, Fe4531 and Ca4227. This choice 
is not unique, but we checked that other index combinations lead to very 
similar results. Considering only these four indices, we compute the 
$\chi^2$ surface as a function of population age and metallicity for 
all nine spectra in our sample. We sample all combinations of age and 
metallicity available in the model, i.e. four metallicities and 220 ages, 
ranging from $10^6$ to $2 \times 10^{10}$ years, roughly logarithmically
spaced. The resulting best fit $\chi^2$ values are quoted in Table 
\ref{t:ind_age}. The $\chi^2$ values in Table \ref{t:ind_age} far exceed 
the degrees of freedom (D.O.F.=2), just another manifestation of the 
systematic mismatch between the SSP models and the data. 
To derive uncertainties, 
we have rescaled the resulting $\chi^2$ to D.O.F. at its minimal value. 
As the coverage of the models in $Z$ is sparse, the uncertainties in 
metallicity are noted as two different possible values in some cases. 
NGC~7418 has two minima that are almost equally good fits (a consequence 
of the double valued H$\delta_A$). The other minimum is at 
log($\tau$) = 6.86$_{-0.06}^{+0.05}$ and $Z$=0.008. The values we quote 
are marginally preferred because the minimum is somewhat 
broader and the age agrees with the one that would be obtained if 
one would marginalize over all metallicities.
The age quoted here also agrees roughly with the age that 
was determined by the spectral fitting (Section \ref{s:SSP}).

%%%%%%%%%%%%%%%%%%%%%%%%%%%%%%%%%%%%%%%%%%%%%%%%%%%%%%%%%%%%%
\subsection{Mean Ages from the Continuous Blue Spectra}
\label{s:SSP}

We now use the full blue SED from 3650 to 4600 {\AA} to 
derive the best fit SSP in a $\chi^2$ sense for each of 
the spectra, via 
\begin{equation}
 \chi^2(\tau,Z,A_I) = \sum_{i=1}^{N_{pix}} \left[ \frac{g_i - t_i(\tau,Z,A_I)}{\sigma_i} \right]^2,
\end{equation}
where $\tau$ is the age of the SSP, 
$g_i$ is the observed SED, $\sigma_i$ is the 
standard deviation on each bin in the SED, $t_i$ is 
the SSP SED and the sum is over all $N_{pix}$ pixel values in this 
spectral range. We sample metallicities at 
$Z$=0.004, 0.008, 0.02, 0.05, where $Z$=0.02 is solar. We have 
chosen 14 SSPs, roughly logarithmically spaced in age, at 
$(1,3,6) \times 10^{(6,7,8,9,10)}$ yrs, plus a last one with 20 
Gyrs. It will be shown later on that this corresponds to the 
age resolution we can actually achieve. We also consider reddened 
model spectra, using the extinction law of Cardelli et al. (1989), which 
was derived for non-starbursting systems. The extinction is quantified 
by the $I$-band extinction A$_I$ (more exactly the 
A($\lambda$) at $\lambda$=8000 {\AA}) which we sampled 
in steps of 0.05 mag (for comparison: the foreground Galactic 
extinction is of the order A$_I$ = 0.1). We 
reference to the $I$-band, because imaging (from B02) and dynamical 
mass to light ratios (from Paper I) exist for this band. 

In this Section and the following, the data were 
degraded to the resolution of the model in exactly the same way as 
in Section \ref{s:lick}. Both model SSPs and 
data were rebinned logarithmically to allow for velocity shifts 
between the model and the observed spectra.

We limited the fit to the wavelength range 3650 to 4600 {\AA} 
for several reasons: the full spectrum covers three CCD chips 
of UVES and even small errors in the relative flux calibration between the 
chips would distort the continuum slope and thus age and extinction 
estimates. Also, the blue part of the spectrum 
has most of the absorption lines useful for analysis. We trim the 
edges of the spectra to eliminate possible edge effects. 

For the calculation of $\chi^2$, we use a code implementing the method 
presented in Rix \& White (1992), which was originally designed for the 
measurement of the internal kinematics of galaxies. 
The software has been used 
and tested in a number of scientific projects (e.g. Rix et al. 1995, 
Sarzi et al. 2005, and Paper I). Although it was originally 
designed to be used for kinematic studies, certain aspects of this 
code make it a useful tool for population synthesis studies. 
In the present section it provides a convenient wrapper 
around a simple $\chi^2$ computation with several advantages: 
first the software automatically matches the position 
of stellar absorption features between templates and objects. 
Second, the software also allows for a convenient way to mask emission 
lines in the spectrum, consequently these do not contribute to the 
finally derived $\chi^2$. We are thus able to minimize 
the influence of the Balmer emission lines on the derived age, in 
stark contrast to methods based on the Lick indices. In our original 
high-resolution spectra, even small emission lines are readily seen 
as spikes in the absorption troughs. In all convolved spectra used in 
this and the following section, we mask the region of the 
[O\textsc{II}] line at 3727{\AA}. We additionally mask the Balmer 
emission lines in five objects, using windows appropriate for the 
resolution of the data. 
For NGC1042 we also mask the [Ne\textsc{III}] emission line at 3868.8{\AA}, 
as it is readily seen in emission. Each of the non-masked 
pixels in the spectrum is weighted with its specific error 
during the fitting process. We now discuss the results for the 
best-fit SSP, with given age, metallicity and extinction, to 
the object spectrum.

Table \ref{t:SSP_age} lists the best fit ages $\tau$, metallicities $Z$ 
and extinctions A$_I$ as well as the $\chi^2$ value for the best fit
N$_{\mbox{DOF}}$ $\approx$ 900. In contrast to what we find for the 
index method, the best $\chi^2$ is here sometimes lower than 
N$_{\mbox{DOF}}$, which possibly indicates that the UVES reduction 
pipeline provides overestimated errors. Table \ref{t:SSP_age} also 
quotes the mass-to-light ratios for the best fitting age from the 
population synthesis model.

\begin{figure}[tbp]
\begin{center}
\includegraphics[angle=0,clip,width=1\hsize]{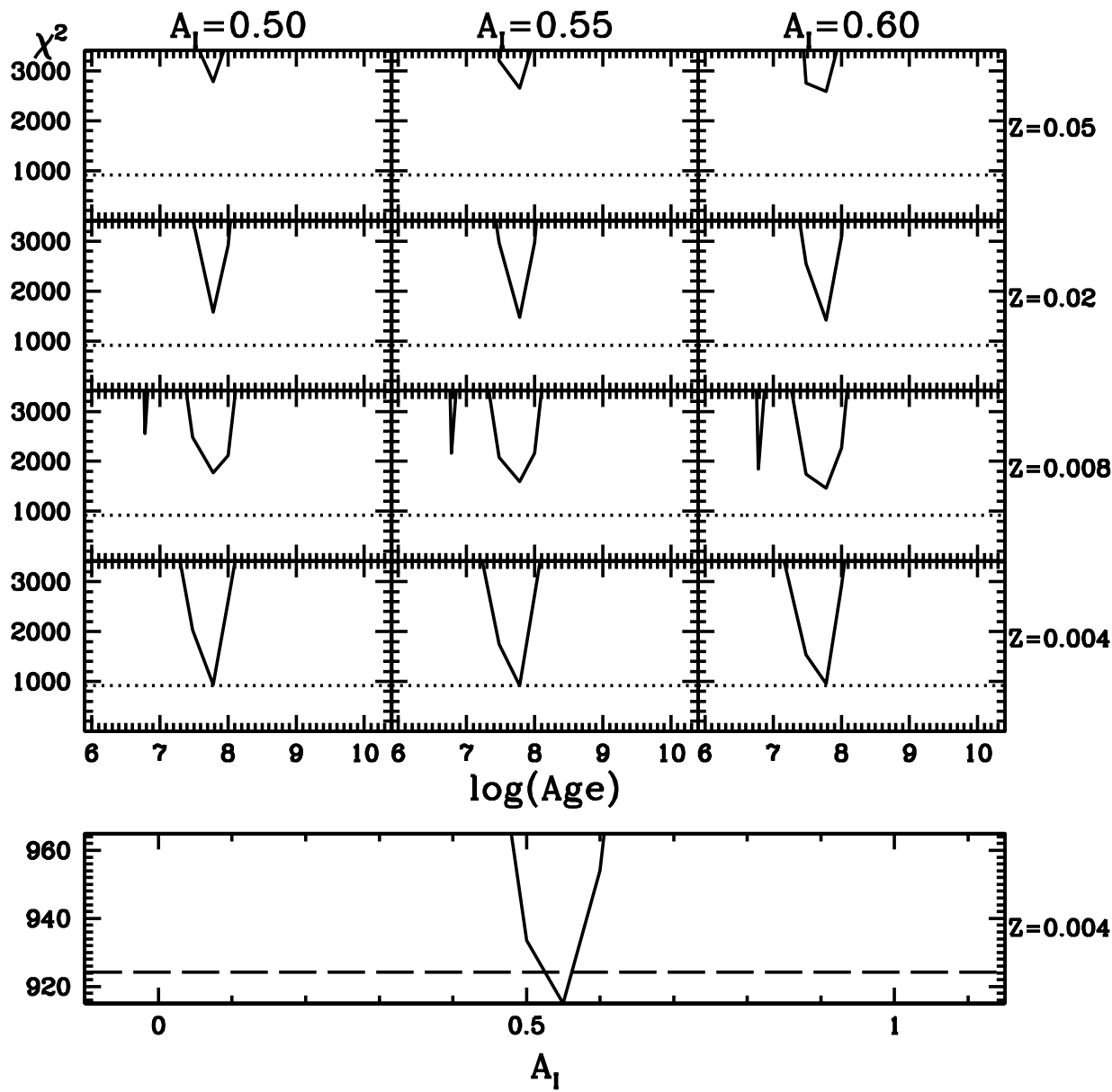}
\caption[all_chi2]{$\chi^2$ as a function of SSPs of different 
age, metallicity $Z$ and extinction A$_I$ for 
the nuclear spectrum of NGC~7418. The upper panel shows $\chi^2$ 
as a function of log(age) for different values of $Z$ and A$_I$. Note that 
the best fit age is independent of the chosen metallicity 
or extinction. The dotted line is the minimum $\chi^2$. 
The lower panel shows in more detail that the 
best fit extinction is well defined in a $\chi^2$ sense. Here, 
we plot $\chi^2$ as a function of A$_I$ for the values of $Z$ and age 
that correspond to the best fit. The dashed line corresponds 
to the 99\% confidence limit.}
\label{f:chi2_7418}
\end{center}
\end{figure}

Representing $\chi^2$ as a function of all three parameters
is difficult. 
We chose to display the full $\chi^2$ parameter space for one 
example spectrum (NGC~7418) in Figure \ref{f:chi2_7418}. It is 
reassuring to note that the best fitting age does not depend
on either metallicity or extinction. The best fit age does 
not change by more than 0.3 dex for a reasonable range of these 
two parameters in any of the spectra we analyze here. To avoid 
the rescaling of $\chi^2$ to N$_{\mbox{DOF}}$ (as done in Section 
\ref{s:lick}), we therefore adopt 0.3 dex as a conservative error 
estimate on the derived ages in Table \ref{t:SSP_age}. Note that 
this justifies in retrospect our choice of age intervals for the 
SSP templates. The SSP metallicity also appears to be well-defined 
quantity as well, 
as its own best fit value changes only for extinction values far 
from the best fit value. We therefore take the error in $Z$ to be 
equal to half the step-size with which we sample the metallicity 
range; i.e., $\Delta Z \approx 0.2$ dex.

The formal uncertainties on the extinction are often lower than the chosen 
step size of 0.05 mag. If the $\chi^2$ contours are rescaled to 
$\chi^2_{\mbox{best}} \approx$ N$_{\mbox{DOF}}$, the formal 99\% 
confidence intervals ($\Delta \chi^2 = 11.3$) are comparable to the 
step size. As we show in Section \ref{s:mixpop}, fitting 
multiple age populations may lead to differing extinction estimates. 
But a detailed analysis of the extinction errors is not crucial 
in the present context, as we are focused on determining the 
population ages. As shown in Figure \ref{f:chi2_7418}, the inferred 
age is quite insensitive to the assumed extinction.

\begin{figure}[tbp]
\begin{center}
\parbox{1\hsize}{
\includegraphics[angle=0,clip,width=1\hsize]{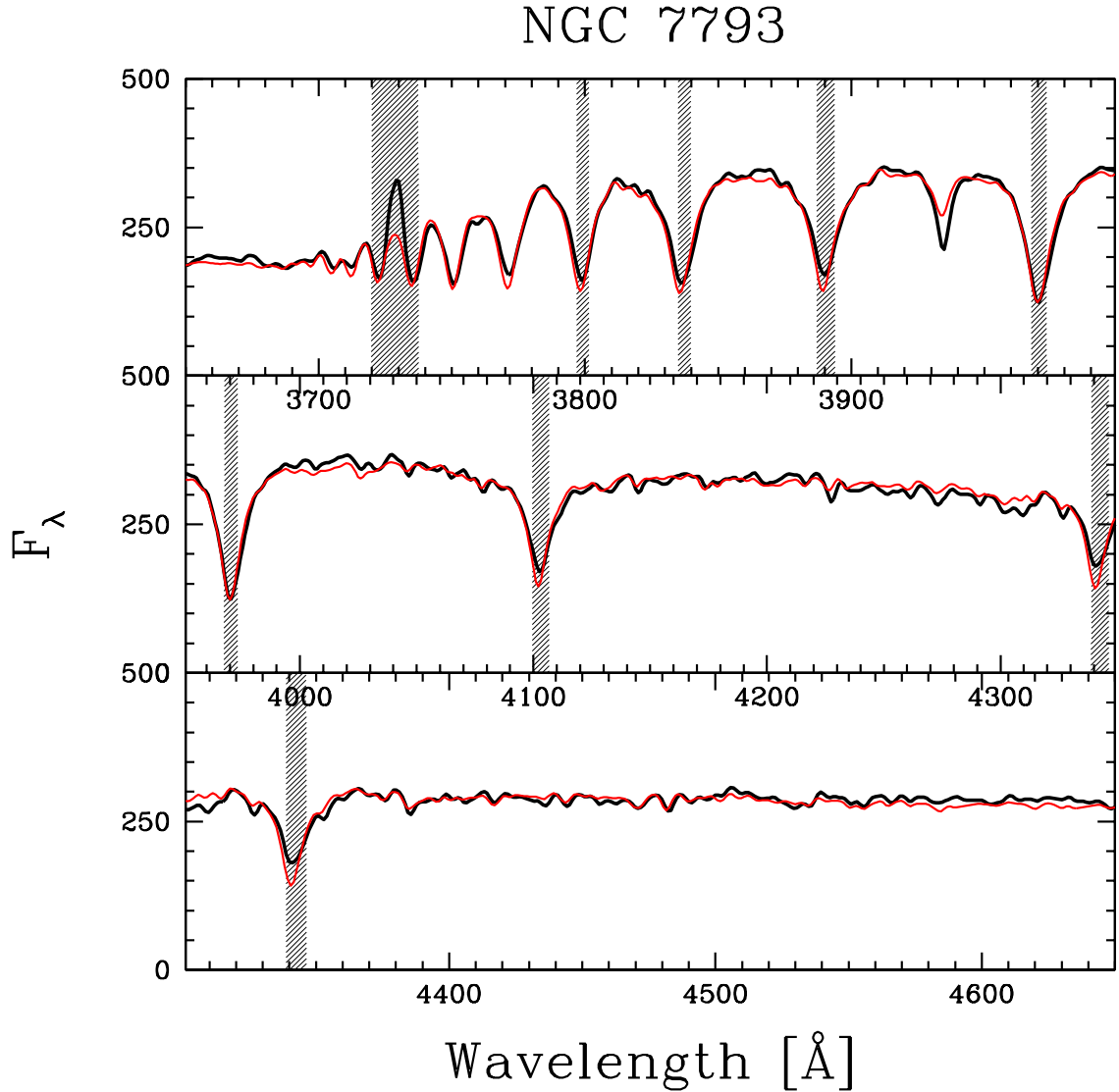}
\caption[ngc7793]{Best single age fit for NGC~7793 (thin dotted 
line, red in the online edition) compared to the observed spectrum 
(thick black line, arbitrary flux scale). Note that the 
Balmer and [O\textsc{II}] emission lines are excluded from the fit
(shaded areas). 
Also note that the single age fit fails to fit the Ca K line 
at 3934 {\AA} and smaller continuum features in the red. 
The panels have a scale with a small amount of overlap in 
wavelength (so the same absorption line is sometimes seen in
multiple panels).}
\label{f:SSP_fits}}
\end{center}
\end{figure}

Figure \ref{f:SSP_fits} shows the observed spectrum of the nuclear 
region in NGC~7793 (which has the highest S/N of all observed objects), 
along with the SED of the best-fit single age population. 
In general, the fit is quite good and reproduces also small 
features of the observed spectrum. As the 
Balmer and [O\textsc{II}] emission lines are excluded from the 
fit, they do not influence the determined age. On the other 
hand, several characteristic features are not fit well. In the case 
of NGC~7793 this includes the Ca K line and the redmost range of 
the spectrum.

In summary, the age of the best fitting single age stellar 
population is robust with respect to the chosen metallicity 
and extinction. The age estimates agree well with those determined 
from the spectral indices, confirming that most clusters are 'young', 
i.e. their age is much smaller than a Hubble time.

%%%%%%%%%%%%%%%%%%%%%%%%%%%%%%%%%%%%%%%%%%%%%%%%%%%%%%%%%%%%%
\subsection{Fitting Composite Stellar Populations}
\label{s:mixpop}

There are different reasons to expect that the NC spectra 
are only poorly described by SSP models. First, some of the spectra 
actually represent an aperture that encompasses cluster light 
and disk light in similar amounts. Their spectra should therefore be better 
described by a mixture of differently aged stellar populations. 
Second, the M/L$_I$ ratios derived from the dynamical modelling 
of Paper I are usually twice as high as those derived from the SSP fits, 
a discrepancy that is larger than the uncertainties. This can be 
interpreted as due to the presence of an older underlying 
population that has higher M/L
and therefore is underrepresented in the observed spectrum. 
The younger population with lower M/L dominates the spectrum, 
though it may contribute only little to the mass. We therefore want 
to test if we can directly detect mixed stellar populations in the 
nuclear spectra. Ideally we would like to date the last burst of 
star formation as well as the oldest population in the spectrum. 

As described at the beginning of Section \ref{s:age}, any composite star 
formation history can be described by a linear superposition of multiple 
bursts of star formation. The task to find the best linear weight for 
each SSP in the template set to fit the total spectrum is implemented 
in our software through the algorithm of Lawson and Hanson (1974). 
We thus generalize the code used in Section \ref{s:SSP} to allow for 
composite populations with multiple ages. We use the same template 
library and the same emission line masks as for the SSP fitting. For 
the sake of a well-defined answer we reduce the number of available 
templates to one mean metallicity for all templates in one set and 
then fit once for each metallicity available. It would also 
be possible that every population, especially the young ones, could 
have a different extinction. We here however assume that one extinction 
is sufficient to describe all populations in each object.

For each object we then explore the $\chi^2$-surface for a wide 
range of metallicity and extinction, deriving the best fitting 
composite age fit at each point. Table \ref{t:mix_age} 
lists the M/L$_I$ and mean luminosity weighted age in the 
$I$-band $\langle \tau_I \rangle$ of the best fit for every object 
together with the associated metallicity 
$Z$ and the extinction A$_I$.

To derive uncertainties on the parameters of interest, 
we follow the same approach as in Section \ref{s:lick}, 
rescaling the errors (or $\chi^2$ surface) in each object to 
yield $\chi^2_{\mbox{best}} \approx N_{\mbox{DOF}} \approx 900$. 
The exact number is different for each spectrum, as different 
regions of the spectrum are clipped because of emission lines. 
Again as in Section \ref{s:lick}, errors projected on a two 
parameter space can then be derived from iso-$\Delta \chi^2$ 
contours. It turns out in all cases that no other value of $Z$ 
yields a fit within $\Delta \chi^2$ = 29 of the best fit, 
which corresponds to the 99\% confidence interval in 14 dimensions. 
We therefore take the uncertainty in $Z$ to be equal to half 
the step-size with which we sample the metallicity 
range; i.e., $\Delta Z \approx 0.2$ dex. For A$_I$ the 
typical uncertainty is of order the step size, i.e. $\Delta \mbox{A}_I 
\approx 0.05$. Table \ref{t:mix_weights} lists the 
mass-to-light ratio (M/L$_I$) and the mass fraction that each 
SSP contributes in the best fit. We here list mass fractions, 
though the fit determines light fractions. Both are related by 
the M/L$_I$ given in the first column of Table \ref{t:mix_weights}. 
The mass weights directly imply a star formation rate over the 
age range that each SSP represents. Figure \ref{f:chiplane} shows 
cuts through the $\chi^2$ surface as a function of $Z$ and A$_I$ for 
two objects, namely NGC 300 and NGC7418.

The derived M/L$_I$ and $\langle \tau \rangle$ depend on a total 
of 16 parameters (14 template weights, metallicity and extinction). 
To derive meaningful error estimates one would have to cover the 
full 16-dimensional $\chi^2$ space. This is computationally too 
cumbersome and hard to interpret. We therefore will give a more 
empirical discussion of the significance of the 
results in the next Section.

%%%%%%%%%%%%%%%%%%%%%%%%%%%%%%%%%%%%%%%%%%%%%%%%%%%%%%%%%%%%%
\subsection{Comparing the Different Approaches}
\label{s:age_methods}

We now compare the three different approaches used to derive mean ages, 
metallicities, extinction parameters and mass-to-light ratios 
for integrated SEDs of star clusters. In this discussion 
we will refer to the approach of Section \ref{s:lick} as the 
index method, to the approach of Section \ref{s:SSP} as 
the SSP method and finally to the approach of Section \ref{s:mixpop} 
as the composite fit.

It is first noteworthy that although the model indices were measured 
on the exact same spectra that are used for the spectral fitting, 
the index method does seem to match significantly less well than 
the spectral fitting in the sense that the reduced $\chi^2$ of the 
best fitting model is in general larger by one order of magnitude. 
This can be mainly attributed to our neglect of templates with 
complicated star formation histories (SFH) when comparing the data 
with the model indices. While the SFHs of galaxies are conventionally 
taken to be decaying exponentials, we have no a priori knowledge 
on the functional form of the SFH of NCs. Additional problems arise 
because the index method purposefully relies on a very small wavelength 
range, to extract specific information from the spectrum - it 
compresses the information. However, a specific disturbance to 
this small wavelength range, e.g. emission lines, can lead to 
significant changes in the reliability of the method. Also the 
metallicity determinations will be affected by the fact that the 
available wavelength range does not grant us access to indices 
which do not depend on metal-abundance ratios. The 
spectral fitting on the other hand does use the full information 
content of the spectra and provides a means to correct for a 
disturbance in a small wavelength range by masking it from the fit
(although it is still affected by the lack of models with varying 
abundance ratios). 
It is therefore the method of choice in the following discussion. 
Spectral fitting, however, does not allow to distinguish between 
the effects of different parameters easily, because the change 
of one parameter does not only affect a single feature in the 
spectrum but rather changes the whole SED. This interpretation 
problem will stay with us during the rest of this Section. 

A comparison of Table \ref{t:ind_age}, \ref{t:SSP_age} and 
\ref{t:mix_age} shows that all three approaches agree and 
measure in general a subsolar metallicity. There are 
some outliers, namely NGC~2139 in the index method and the composite 
fit, and NGC~428 in the SSP fit. However, the case of NGC~7424 
(see Figure \ref{f:chiplane}) shows that our metallicity results 
may not be very robust for every single object. 
We therefore conclude that while we have evidence that the nuclei of 
bulge-less spirals have on average a slightly sub-solar metallicity 
($\langle Z \rangle = 0.015 \pm 0.004$), the robustness of the 
metallicity measure for single objects remains uncertain. 
When looking for comparison values in the literature, 
we find the following values for NGC~300. Butler et al. (2004)
measured a metallicity of 0.006 from color-magnitude 
diagrams for the stars in the disk of NGC~300. Individual 
stars in NGC~300 have also been targeted spectroscopically. 
Urbaneja et al. (2003) have measured metallicities of 0.006
and 0.02 for two stars in the outskirts and in the center of 
the galaxy respectively. Zaritsky, Kennicutt \& Huchra (1994) 
have determined Oxygen 
abundances for H\textsc{II} regions in NGC~300 and NGC~7793. 
Assuming solar abundance ratios, we transfer these values to 
a value of $Z$, using
\begin{equation}
\log(\mbox{Z}) = \log \mbox{Z}_{\odot} + [12 + \log(O/H)]_{\mbox{gal}} - [12 + \log(O/H)]_{\odot}, 
\end{equation}
where we take $[12 + \log(O/H)]_{\odot} = 8.93$ and $Z_{\odot} =
0.019$ (Anders \& Grevesse, 1989). We thus obtain $Z = 0.012
\pm 0.01$ at r = 2.5 kpc for NGC300 and $Z = 0.015 \pm 0.03$ at r =
1.7 kpc for NGC7793. The $Z$ values we derive for our NCs are almost 
a factor
of two lower.  There might be several physical reasons for this
discrepancy, as e.g. population gradients, differences between
gas-phase and stellar metallicities, or non-solar abundance ratios, 
but the largest issue remains the systematic uncertainties 
of the various measurements.

Concerning extinction, it should be noted first that the 
continuum slope itself is a function of age and 
metallicity. Therefore, there is a trade-off 
between including an old population in the composite fit 
and applying a higher extinction to an SSP spectrum. 
While extinctions measured 
from SSPs therefore are almost certainly 
biased, the extinctions measured from the composite fit 
should be less influenced by systematic biases. 
For our data, an additional significant caveat exists in that we 
measure extinction over a very limited wavelength range. 
While age is quite insensitive to the derived extinction 
on such a small wavelength range (compare Section \ref{s:SSP}), 
the value of the extinction itself may not be well constrained.
Extinctions inferred from the composite fit are generally lower
than those from the SSP method, probably implying that in the 
SSP case the contribution from an old stellar population is 
substituted by the higher extinction correction. 
Of the two objects with highest extinctions, NGC~7418 
also has a dust lane that is visible on the HST $I$-band image 
from B02. While the NC in NGC~428 is rather old and 
we would therefore expect it not to be surrounded by a lot of gas, 
the galaxy  NGC~428 as a whole probably is a 
recent merger and is forming stars at a rather high rate (Smoker 
et al. 1996), implying a gas and dust reservoir to 
be present in this galaxy. The spectral indices 
shown in Figure \ref{f:metage} imply a significant amount 
of extinction for this object. Fairly low internal extinction 
has generally been measured in extreme late-types spirals 
from broad band colors (e.g. de Blok et al. 1995; Tully et al. 1998) 
and from spectroscopy (e.g. Roennback \& Bergvall 1995). 
The last authors find a median A$_V$ = 0.3 (A$_I \approx$ 0.2) 
for their sample of blue low-surface brightness galaxies. 
Urbaneja et al. (2003) also determined extinctions for the two 
aforementioned single A giants in NGC~300 and found that 
``magnitudes and colors are consistent with almost no reddening 
for both stars''. While these literature results are not 
contradicting our findings, it should be noted 
that there is no reason why the nucleus should not 
have an extinction value of its own, possibly different 
from the rest of the galaxy. 

Concerning age, the index method and the SSP method agree 
remarkably well within the uncertainties - despite all problems with 
the former. Thus, both suggest a distinction into two groups 
of nuclei: those that are $\lesssim 0.1$ Gyr (NGC~1042, NGC~1493, NGC~2139, 
NGC~7418, NGC~7424, NGC~7793) and those that are  $\gtrsim 1$ Gyr 
(NGC~300, NGC~428, NGC~3423). Note that none of the 
inferred ages are nearly as high as those of the globular 
clusters of the Milky Way. Looking at the literature, Diaz 
et al. (1982) found that the spectrum of the nucleus of 
NGC~7793 is dominated by A or early F stars near the main 
sequence, in agreement with our findings.
The mean luminosity-weighted age as derived in the $I$-band 
from the composite fit is older than the age inferred from the SSP 
fit in all nine cases. The division into 3 old and 6 
young nuclei is however still valid in the sense that NGC~300, 
NGC~428 and NGC~3423 all have less than 3\% of their 
mass in any population younger than 1 Gyr. Additionally, NGC~1042 
needs to be added to this category. This is a particularly striking 
example of the trade-off between extinction and old population. 
For the 5 other spectra, the mass fraction of young populations 
is at least 10\%. 
Figure \ref{f:SC} shows the different extinctions and mean 
luminosity weighted ages we obtain with the SSP method and 
the composite fit. This plot exhibits clearly the shift from young 
to old mean age that occurs when comparing the SSP to the composite fit. 

\begin{figure}[tbp]
\begin{center}
\parbox{1\hsize}{
\includegraphics[angle=0,clip,width=1\hsize]{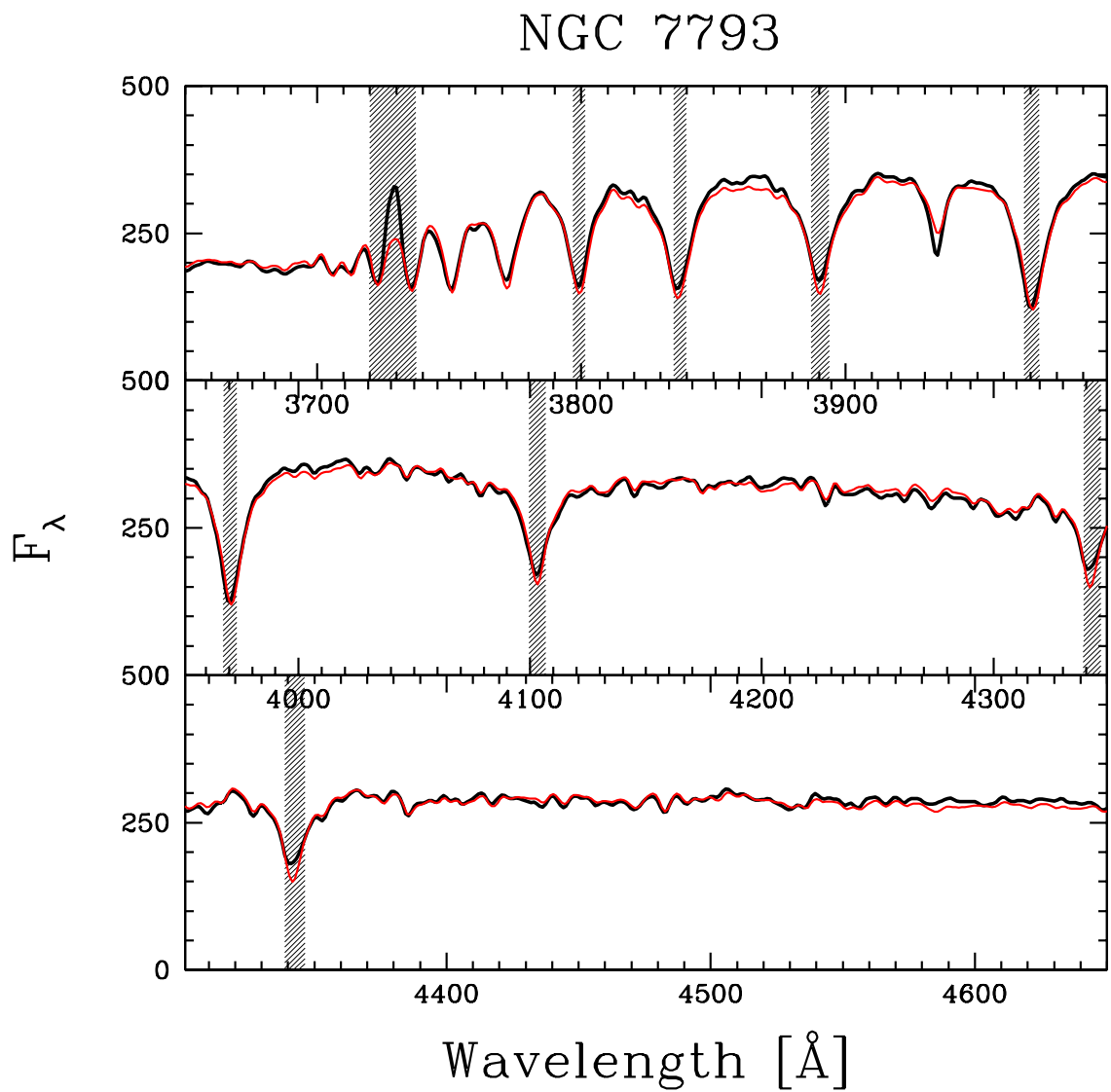}
\caption[ngc7793]{Best composite fit for NGC~7793 (thin, dotted line, 
red in the online edition) compared to the observed spectrum (thick line, 
arbitrary flux scale). Note that the Balmer and [O\textsc{II}] 
emission lines are excluded from the fit (shaded areas).
Also note that the fit to the Ca K line at 3934 {\AA} and the 
continuum in the red 
is improved, as compared to Figure \ref{f:SSP_fits}. 
Note that the panels have a scale with a small amount of overlap in
wavelength (so the same absorption line is sometimes seen in
multiple panels).}
\label{f:mix_fits}}
\end{center}
\end{figure}

\begin{figure}[tbp]
\begin{center}
\parbox{1\hsize}{
\includegraphics[angle=0,clip,width=0.99\hsize]{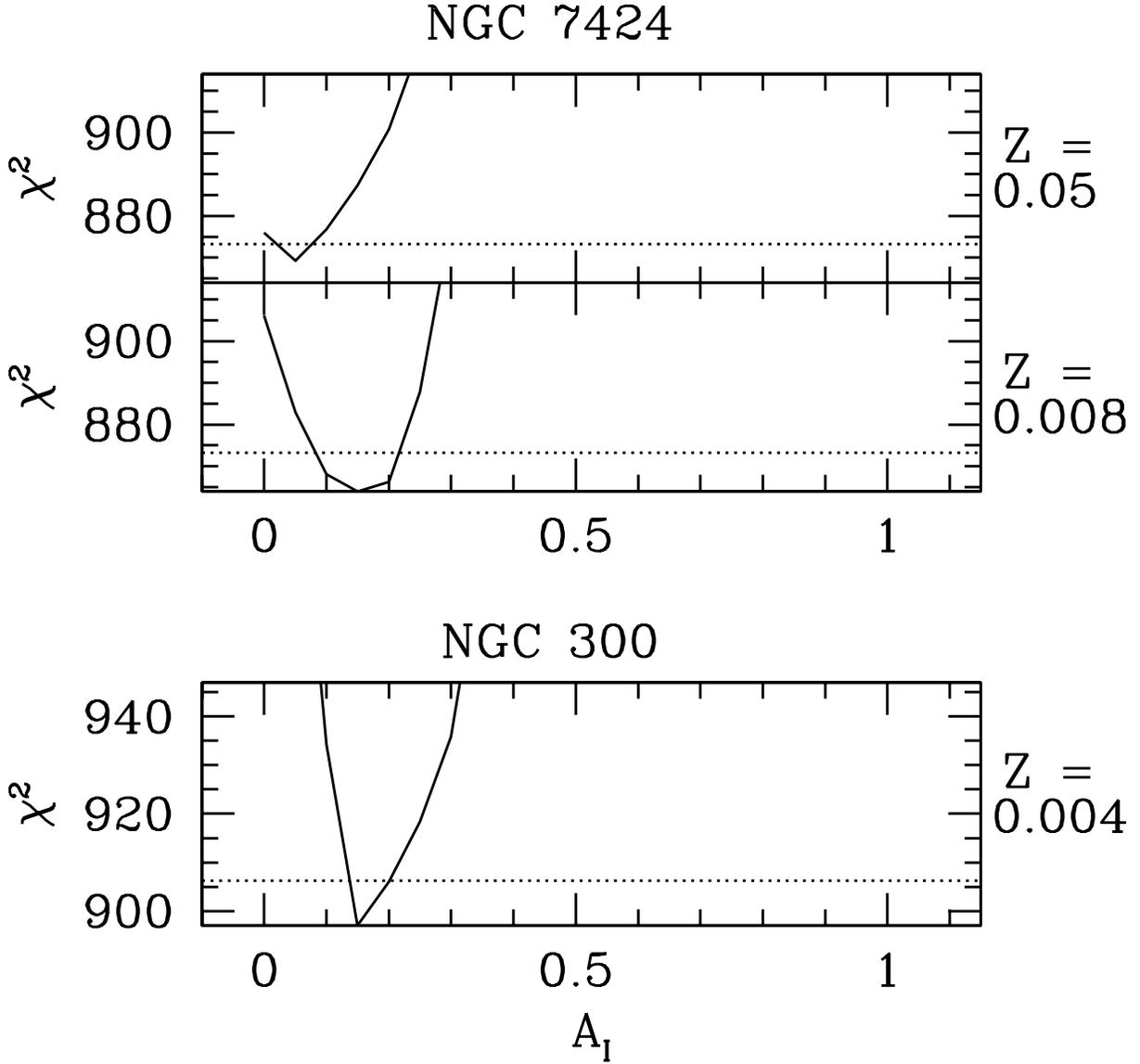}
\caption[chiplane]{Cuts through the $\chi^2$-surface as a function 
of metallicity and extinction for two objects, obtained by optimizing 
the possible composite age distribution at each point. Panels that would 
be empty because of high $\chi^2$ values have been omitted (compare Figure 
\ref{f:chi2_7418}). 
Note that both quantities are well defined from the point of view of 
$\chi^2$ statistics for most objects as for NGC~300. The case of 
NGC~7424 however shows that metallicity might not be very robust for 
this method. The $\chi^2$ value of the best fit has been 
adjusted to correspond to the number of free parameters. The dotted 
line represents the 99\% confidence interval in 14 dimensions, 
$\Delta \chi^2 < 29$.}
\label{f:chiplane}}
\end{center}
\end{figure}

\begin{figure}[tbp]
\begin{center}
\parbox{1\hsize}{
\includegraphics[angle=0,clip,width=0.90\hsize]{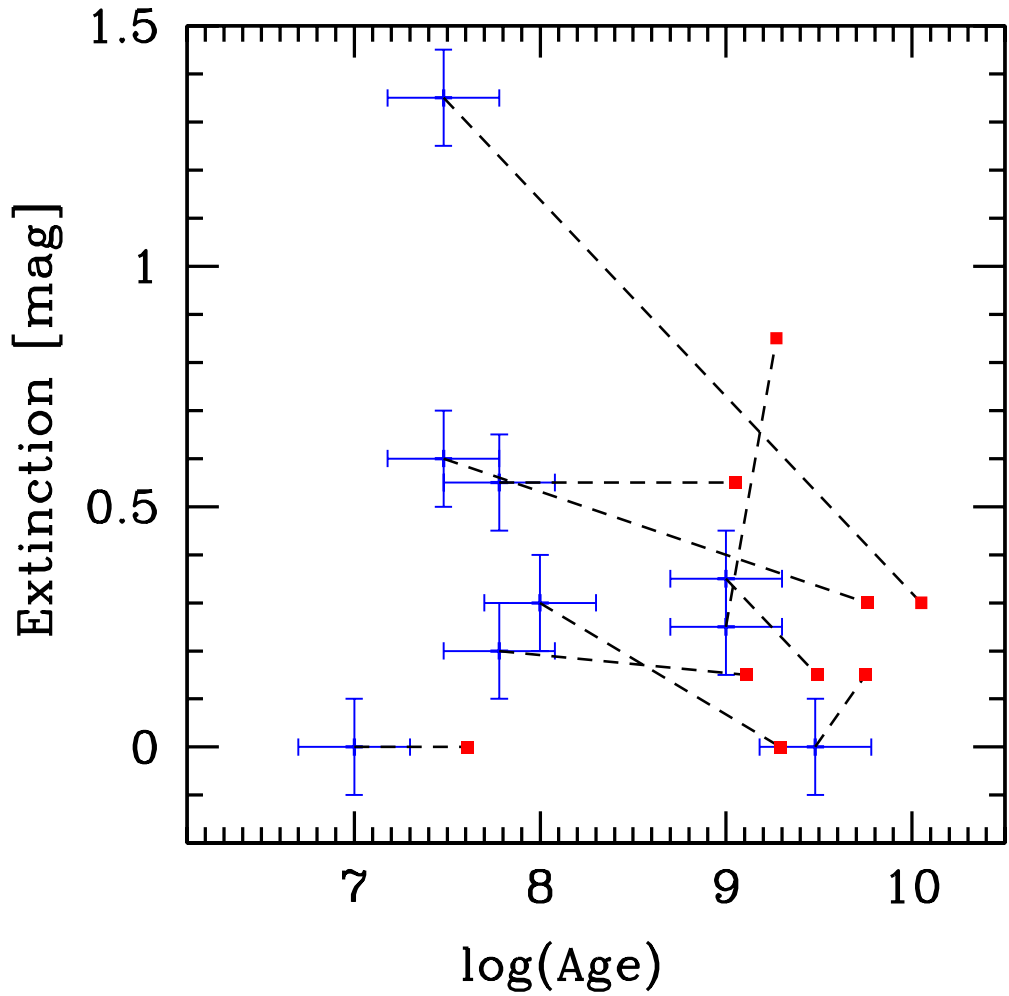}
\caption[SC]{Comparison of the best fit values for the luminosity 
weighted mean age and extinction, as derived from the SSP method 
(plus sign with error bars - blue in the online edition) 
and the composite fit (filled square - red in the online 
edition). For each object the two symbols 
are joined with a dashed line.}
\label{f:SC}}
\end{center}
\end{figure}

Compared to the composite fit, only one of the SSP fits is 
statistically acceptable. Although this is not obvious from the 
example shown in Figures \ref{f:SSP_fits} 
and \ref{f:mix_fits}, $\chi^2$ increases by 40 \% from the 
composite fit to the SSP fit on average over all nine 
spectra. A close inspection of the fits backs up these numbers. 
For example in NGC~7793 the fit to the Ca K line as well 
as to the reddest part of the spectrum is visibly improved 
in the composite fits. 
The one object where both $\chi^2$'s are comparable 
is NGC~2139, where a recent, strong burst of star formation 
probably swamps any light from older stellar populations. 

\begin{figure}[t]
\parbox{1\hsize}{
\includegraphics[width=0.49\hsize]{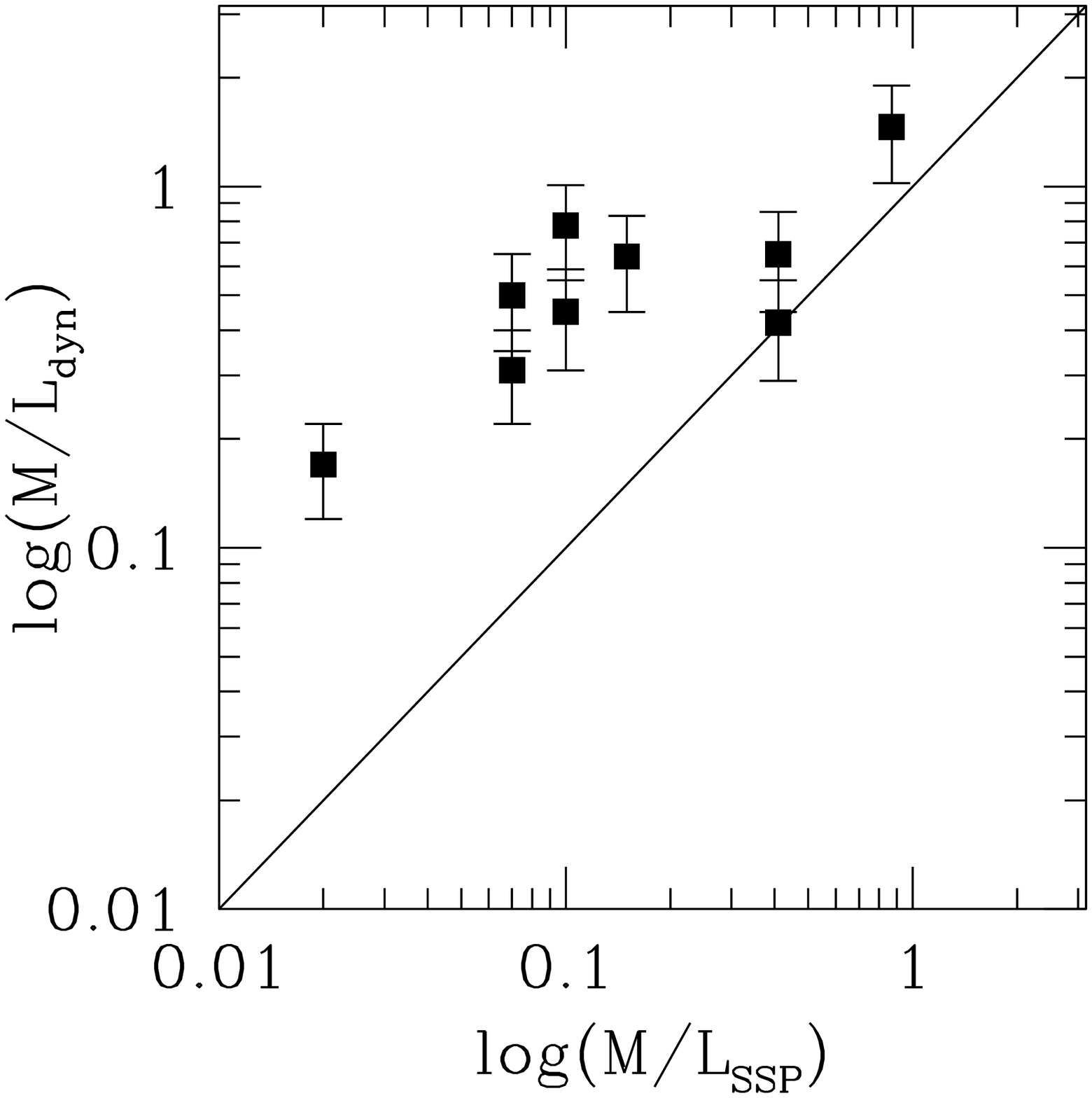}
\includegraphics[width=0.49\hsize]{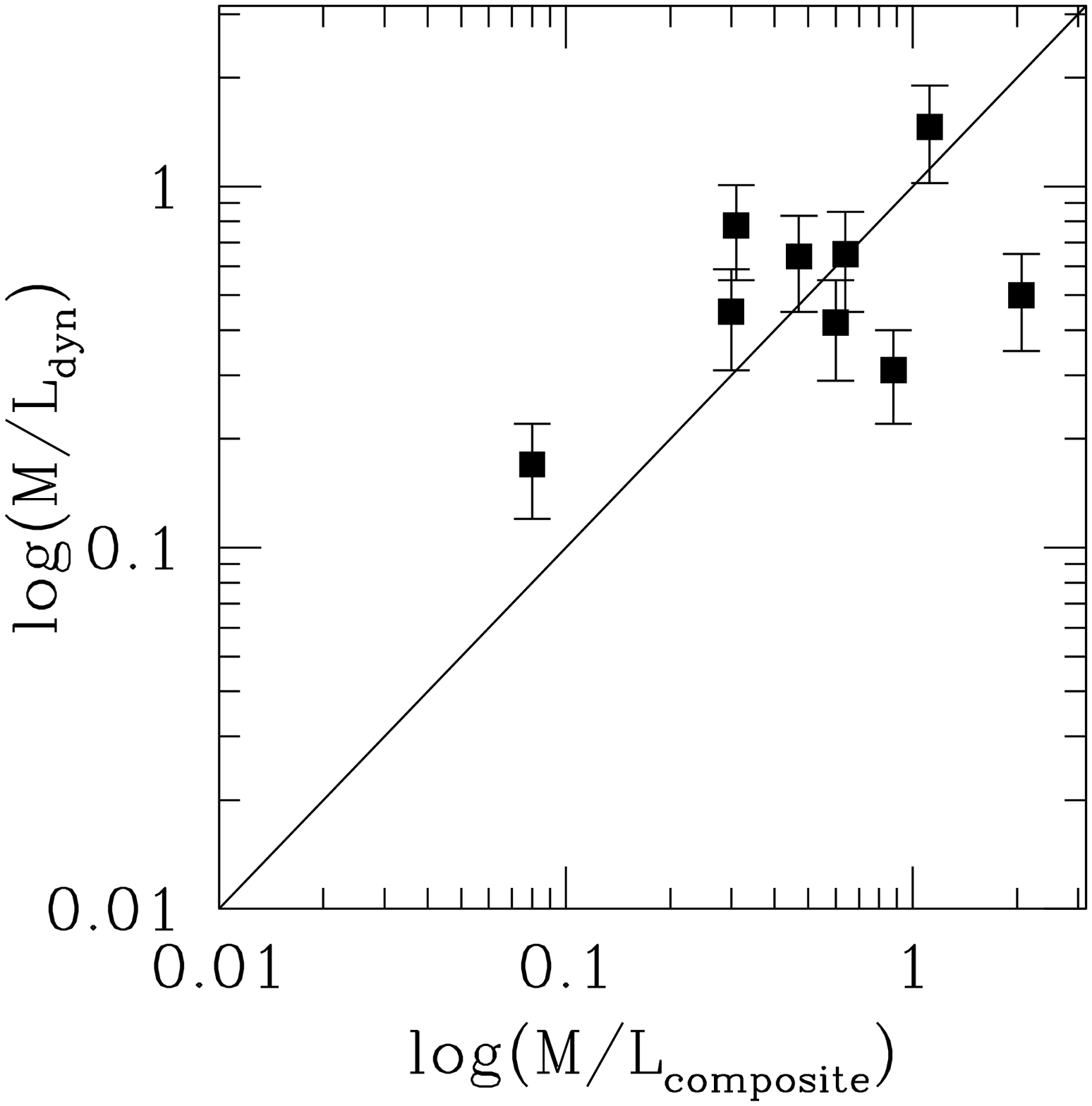}}
\caption{Mass-to-light (M/L$_I$) ratios as derived by the dynamical 
analysis of Paper I against the M/L$_I$ as derived 
from fitting SSPs to the blue spectra (left, Section \ref{s:SSP}) 
or from fitting composite stellar populations (right, Section 
\ref{s:mixpop}). The line indicates a one-to-one relation. Note that 
the M/L$_I$ values of the composite fit are a better match to the 
dynamically measured M/L$_I$ than those from the SSP method. }
\label{f:LtoM}
\end{figure}

However, there is yet another 
indication that the spectra are better represented by 
the composite fit. It was discussed 
earlier that the M/L$_I$ ratios derived from the dynamical 
analysis should be matched by the population analysis. 
Figure \ref{f:LtoM} compares the mass-to-light ratios derived from 
the dynamical analysis to the mass-to-light ratios derived 
from the spectrum fits. The left panel shows the results 
obtained assuming a single-age population in the spectra. 
It is obvious that all but one object lie above the one to one 
relation, indicated as a straight line. This systematic offset 
(on average $\Delta$ M/L = 0.35 with an RMS scatter of 0.20) is reduced 
significantly if the mass-to-light ratio from the composite fit 
is plotted on the same graph (right, on average $\Delta$ M/L = -0.1) --- 
albeit with a large RMS scatter of 0.58. Leaving out the 'outlier' 
NGC~1042, these numbers come down to  $\Delta$ M/L = 0.05 with a 
scatter of 0.28. Note here that this 
comparison is fully justified also in respect of the 
contamination from non-cluster light, because both 
values have been determined on spectra that were extracted 
on the same aperture.

%%%%%%%%%%%%%%%%%%%%%%%%%%%%%%%%%%%%%%%%%%%%%%%%%%%%%%%%%%%%%
\subsection{The Youngest Population Component}
\label{s:lastburst}

\begin{figure}[htb]
\begin{center}
\parbox{0.7\hsize}{
\includegraphics[width=1\hsize]{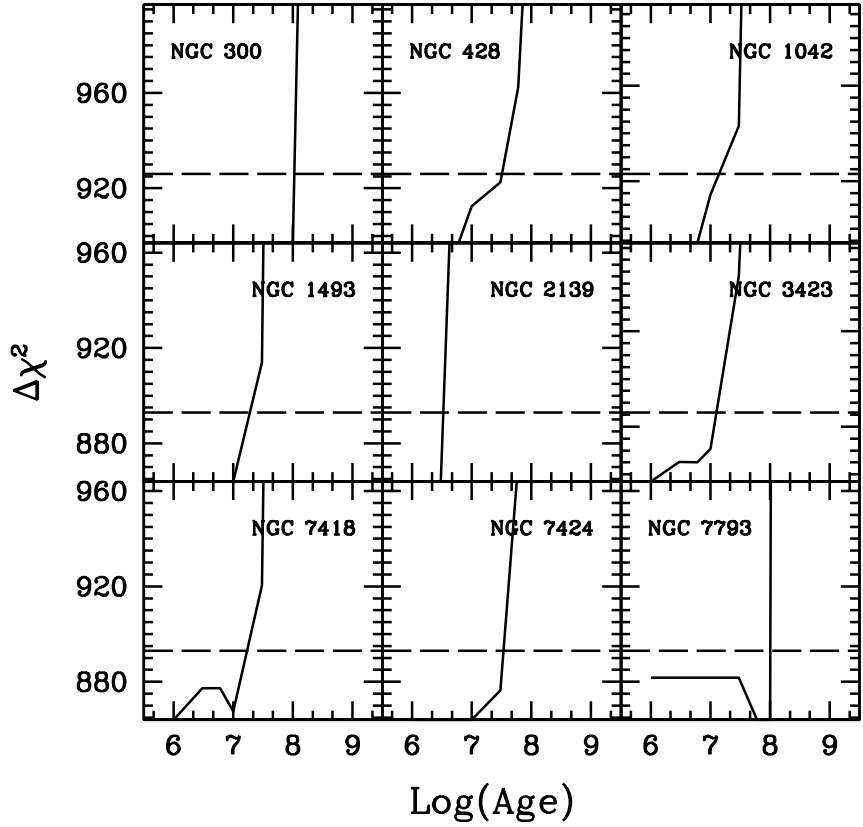}}
\caption{$\Delta \chi^2$ as a function of the age of the youngest 
SSP in the template library for the composite fit. The dashed line 
shows the 99\% confidence region in a 14 dimensional parameter 
space ($\Delta \chi^2 < 29$).}
\label{f:restricted}
\end{center}
\end{figure}

Given the extensive set of 14 templates we fit to each spectrum in 
Section \ref{s:mixpop}, it is difficult to fully explore the topology 
of the$\Delta \chi^2$ contours that encompass the acceptable solutions. 
Instead of exploring all co-variances, we 
will here ask a simpler question: 
how significant is the age of the youngest stellar 
population that contributes to the fit? More specifically, we explore 
the upper age bound on the youngest population component clearly 
required by the data. We derive this maximal age by subsequently 
eliminating more of the young components from the set of templates, 
re-fitting the data at each step. That is, we 
start from the best fit, obtained from a total of 14 template SSPs 
ranging in age from 1 Myr to 20 Gyr. We then omit first the 
youngest SSP age, leaving only 13 templates ranging from 3 Myr on. 
Then we repeat the fit omitting also the 3 Myr template, leaving 
only 12 templates from 6 Myr to 20 Gyr and so on. The maximum age 
of the youngest population age component is then defined as 
the age of the SSP whose 
omission increases $\chi^2$ significantly. The results of 
this procedure are shown in Figure \ref{f:restricted} as 
$\Delta \chi^2$ plots. Although an exact definition of a confidence 
interval is non-trivial here (because of the change of degrees of 
freedom per fit), it is clear for all spectra which SSP's omission 
changes $\Delta \chi^2$ drastically. Note that the fits cannot 
discriminate the age of the youngest population for quite a large 
interval in age in some cases! A conservative estimate 
of the last star formation age is given by the 99\% confidence 
region in a 14 dimensional parameter space ($\Delta \chi^2 < 29$). 
Ages of the last SF burst ($\tau_{lb}$) derived this way are given 
in Table \ref{t:prop}. Across the sample the mean age of the 
last star formation burst is then 34 Myr.

To derive a best estimate and a confidence region on the mass 
of this burst, one could think of varying the mass fraction 
of the last star formation burst. Technically, this would be 
done by subtracting a fraction of the light in the form of 
the SSP with the age derived for the last burst. However, 
we have seen above that the total mass of the population 
fit is subject to considerable scatter. This leads to a 
problem in the present case. Fitting after subtraction 
of a certain SSP may change the total M/L and thus 
$M_{\mbox{NC}}$ erratically.
Indeed, increasing the fraction of light that a specified 
SSP contributes to the total blue luminosity of the object 
spectrum, does not necessarily imply increasing its mass 
fraction too. In the case of an increase of the total M/L, 
the mass fraction of the specified SSP may actually decrease. 
It turns out in practice that this is exactly what happens. 
Of course one can take at face value the mass fractions of 
the ``last fitting model'', i.e., in the sequence of models 
as described above, the model that still includes the SSP 
with the age of the last burst. This model may be 
different from the best fit of Section \ref{s:mixpop}. The 
mass fractions of the youngest SSP in that ``last fitting model'', 
multiplied by the masses as derived in Paper I then yield 
mass estimates for the last burst -- which should be regarded 
as indicative only. In order of ascending NGC number, we find 
the following log(M/M$_{\odot}$): NGC~300:4.5 - NGC~428:4.4 - 
NGC~1042:3.9 - NGC~1493:4.0 - NGC~2139:4.8 - NGC~3423:3.7 - 
NGC~7418:6.2 - NGC~7424:5.1 - NGC~7793:6.1 
(mean log(M/M$_{\odot}$) = 5.5).

We applied a similar technique to test if we can derive 
meaningful results about the contribution of old 
SSPs to the fits. The most notable result is that 
in NGC~2139 even a modest contribution of 10\% 
of the blue light from any population $\ge$ 0.1 Gyr 
is ruled out, in qualitative agreement with the low 
M/L$_I$ derived dynamically. It is quite possible 
therefore that NGC~2139 contains a NC that is in the 
process of forming. However, it should be kept in mind 
that the light of this cluster is dominated by a very 
bright and young population with a low $M/L$. The best-fit 
SSP age is the youngest in our sample ($\log \tau = 6.78$). 
Therefore, in this cluster it is particularly hard to detect 
an underlying older population. Even if an old 
(say $10^{10}$ yr) population contributed only
one percent of the light in the cluster, it would still
contribute as much mass as the young population.
It is also striking that the NC in NGC~2139 is the only 
one, where we measure a super-solar metallicity. Taking this 
result at face value and admitting the cluster was forming 
right now, would require an unusually high metallicity for the 
gas that falls onto the cluster to form the stars.

%%%%%%%%%%%%%%%%%%%%%%%%%%%%%%%%%%%%%%%%%%%%%%%%%%%%%%%%%%%%%
\subsection{How Important is the Disk Contamination?}
\label{s:contamination}

\begin{figure}[hbt]
\begin{center}
\parbox{0.7\hsize}{\includegraphics[angle=0,clip,width=1\hsize]{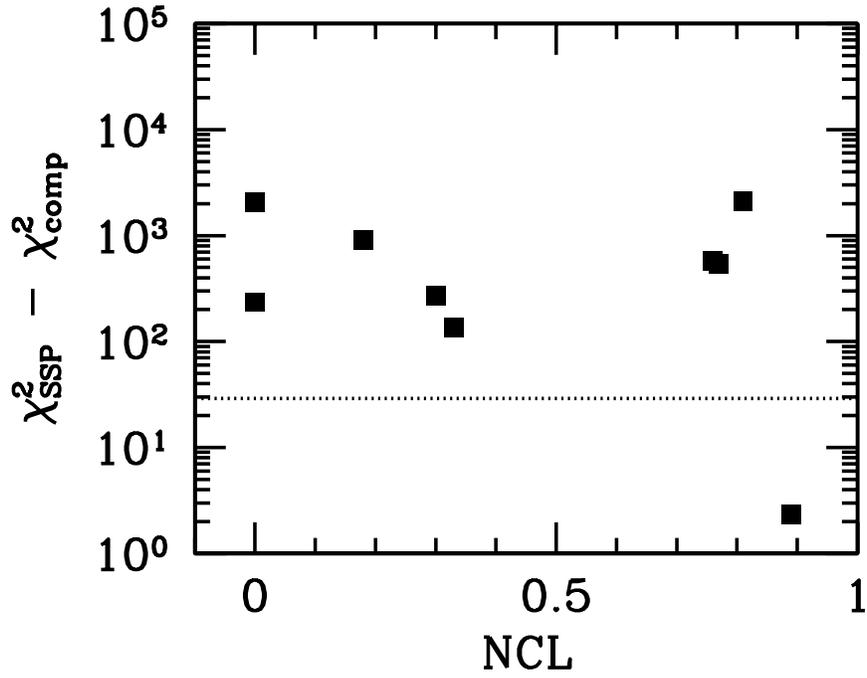}}
\caption[NCL]{$\Delta \chi^2$ between SSP fit and composite fit against 
the fraction of non-cluster light, NCL, in the spectra. The absence of a 
correlation shows that the presence of composite populations is a 
general feature not only of these galaxy centers, but also of the 
NCs.}
\label{f:NCL}
\end{center}
\end{figure}

Strictly speaking, we have up to now only shown that our observed 
spectra are generally best fit by multi-age populations. However, 
in view of the sometimes significant contribution by the light 
of the surrounding galaxy to the observed spectrum, we would like 
to know whether this is true for the NCs as well. 
In other words, is it possible that the NCs
actually do have single-age populations, as is generally true for
e.g. globular clusters and so-called super star clusters? If so,
then our results can be explained only by the fact that we are
observing a mixture of NC and disk light, and either of the
following: (a) the NC and disk have different populations; or (b)
the disk itself has a multi-age population, whereas the NC does
not. In either case we would expect the spectra to be better
described by an SSP in cases where the fraction of non-cluster-light
(NCL) contamination is lower. 
We quantify the quality-of-fit difference between the 
SSP and the composite fit, namely 
\begin{equation}
 \Delta \chi^2 = (\chi^2_{SSP} - \chi^2_{comp}) \frac{N_{pix} - N_{free}}{\chi^2_{comp}}, 
\end{equation}
where $N_{DoF}$ is the number of degrees of 
freedom and $N_{free}$ is the number of free parameters. 
In Figure \ref{f:NCL} we plot this $\Delta \chi^2$ against the 
flux fraction of NCL from Table \ref{t:UVES}. This shows that 
less-disk-contaminated spectra are not more consistent with 
a single-age model. For comparison, the 99\% confidence limit 
in a 14 dimensional $\chi^2$ space is shown as a dotted line. 
This test indicates that the presence of multi-age populations is not
merely the result of disk-light contamination. 
We therefore conclude at this stage 
that the presence of mixed populations is a general feature not 
only of these galaxy centers, but also of the NCs. Of course this 
does not necessarily mean that the populations of the NC and the 
disk are the same. 
We further address this question in Section \ref{s:STIS}.

%%%%%%%%%%%%%%%%%%%%%%%%%%%%%%%%%%%%%%%%%%%%%%%%%%%%%%%%%%%%%
\section{Comparison VLT/UVES vs. HST/STIS Spectra}
\label{s:STIS}

Our population results describe the average stellar
properties within an area of size $\sim 1''$, which is the spatial
resolution of our VLT data. However, the NCs that we are
interested in typically have effective radii $\leq 0\farcs2$ (B04). 
As a result, in many of our spectra most of the observed
light actually comes from the disk, and not from the cluster
(Table \ref{t:UVES}). It is therefore possible that the population 
properties of the NCs are different from the average properties
derived from the VLT spectra. To investigate this issue we also
analyzed spectra obtained at higher spatial resolution. Such spectra
are available for 4 of the 9 clusters in our sample from the Space
Telescope Imaging Spectrograph (STIS) on the Hubble Space
Telescope (HST). These spectra form part of our STIS spectroscopic
survey of 40 NCs in both late-type and early-type
spiral galaxies (HST programs GO-9070 and GO-9783;
P.I. T.~B\"oker). Preliminary results of this work for a subset of the
sample were presented in B\"oker et al. (2003); a final analysis of the
complete sample is presented in a separate paper (Rossa et al. 2006). 

The STIS spectra were obtained with the G430L grating. They cover the
wavelength range 2888.6--5703.2\,{\AA} with a pixel scale of
$2.73$\,{\AA}. The spectra were obtained with a $0\farcs2$-wide slit
({\tt 52x0.2}). The FWHM of the resulting line-spread function is
$1.4$ pixels for a point source and $4.0$ pixels for a constant
surface brightness extended source (Kim Quijano et al. 2003). The
NCs are generally barely resolved at HST resolution,
and fall between these extremes. For a $2.5$ pixel FWHM at 4000\,{\AA}
the spectral resolution is $R = 586$. The spatial pixel scale of the
detector is $0\farcs05$ per pixel. We extracted and co-added the central
four rows from the pipeline reduced two-dimensional spectra. This
yields for each galaxy a one-dimensional spectrum for a $0\farcs2\times
0\farcs2$ square aperture centered on the NC. This
translates to $13.5$ pc $\times 13.5 $pc at the mean distance to the
galaxies. As done previously for the VLT/UVES spectra, an estimate 
for the (small) remaining contribution from the galaxy
disk was subtracted from the STIS spectra.

Both the spectral resolution and the signal-to-noise ratio of the STIS
spectra are worse than for the VLT spectra. Therefore, we do not
repeat here for the STIS spectra the detailed analysis of
metallicities and extinctions previously described for the VLT
spectra. Instead, we kept the metallicity and extinction for each
galaxy fixed to the previously derived values. We then used the exact 
same set of SSP templates of different ages to determine both the 
best single-age fit
$\tau$, and the luminosity-weighted average age $\langle \tau
\rangle$ (in the $I$-band) for the best composite-age population 
fit. For this analysis
we resampled both the template and galaxy spectra to a common
logarithmic scale of $138$ km~s$^{-1}$ per pixel. We then determined the
best-fitting redshift, dispersion, and population for each spectrum.
The dispersion is a measure of the STIS instrumental resolution, and
was found to be in the expected range for all galaxies analyzed. We
used the wavelength range from 3539.7\,{\AA} to 5681.9\,{\AA} for the
analysis. This extends further redward than the spectra used for the
VLT analysis. For practical reasons we used independent 
software for the analysis of the STIS spectra, developed by one of us 
(RvdM), which differs in implementation from that
described previously and used for the VLT spectra. This software is 
based on the van der Marel (1994) pixel fitting routine. Both codes 
solve the same $\chi^2$ minimization problem and we verified through
extensive tests that both codes give consistent answers when applied
to the same problem.

\begin{figure}[hbt]
\begin{center}
\parbox{0.8\hsize}{
\includegraphics[width=1\hsize,height=0.5\vsize]{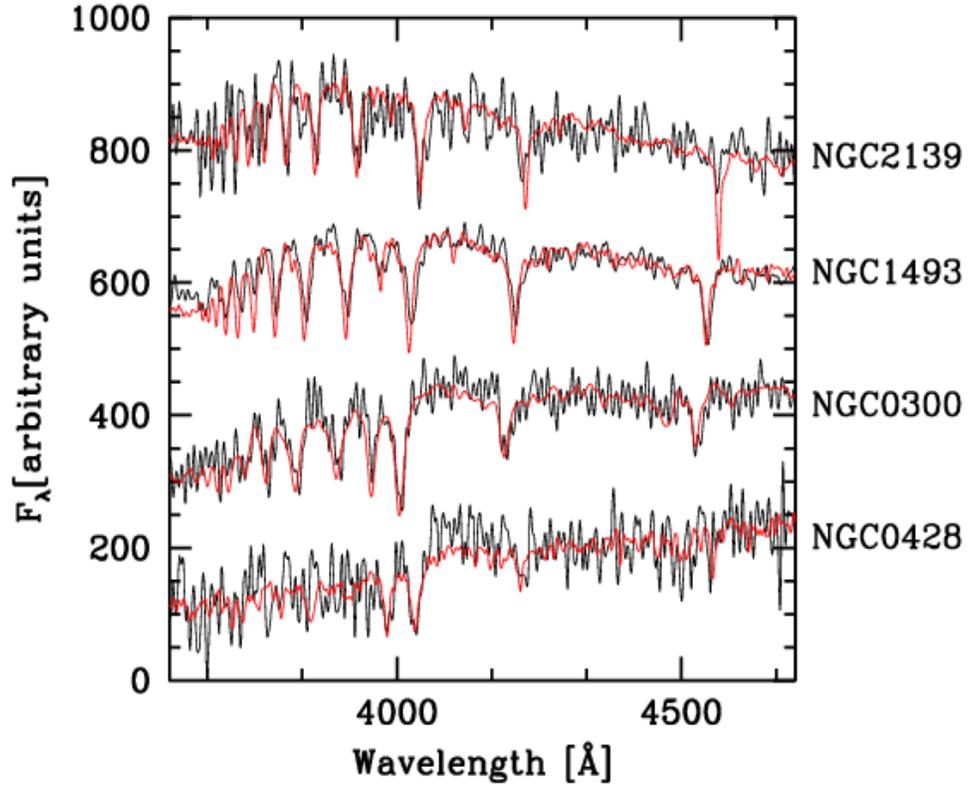}
}
\end{center}
\caption[]{STIS spectra (black) of four NCs in the 
range from 3600\,{\AA} to 4700\,{\AA} overlaid with the VLT 
spectra (red). Note that despite the different apertures (0.2 $\arcsec$ 
vs. 1 $\arcsec$ respectively) the spectra are clearly similar. The spectra 
have been shifted arbitrarily in flux scale.}
\label{f:STIS}
\end{figure}

Figure \ref{f:STIS} shows the STIS spectra and, overplotted, 
the VLT spectra. For this direct comparison the spectra were 
adjusted with respect to velocity (wavelength shift) and 
pseudoflux level. The good agreement between the spectra 
from totally different instruments also strengthens our 
confidence in the response correction we performed for both 
instruments. The composite age population fits for the STIS data 
are similarly good as for the VLT spectra. Table \ref{t:STIS}
compares the age results to those derived from the VLT spectra. 

On average, the results for $\log_{10} \langle \tau \rangle$, the
luminosity weighted mean age, are larger for the VLT spectra by
$0.43 \pm 0.51$ dex in the mean over the sample. For the simple SSP
fit, $\log_{10} \tau$ is again larger for the VLT data than for the
STIS data, in this case by $0.31 \pm 0.11$ dex. These differences
are comparable to the estimated uncertainty in the ages inferred
from the SSP fits to the UVES spectra. They are much less than the
typical age spread between the clusters. We do note that for 
NGC~1493 and NGC~2139 the VLT and STIS spectra give substantially
different results for $\log_{10} \langle \tau \rangle$. This is not
because the spectra or the fits are very different. Instead, the
reason is that for spectra that are dominated by young light, the
luminosity-weighted mean age $\langle \tau \rangle$ is very
sensitive to even a small fraction of the light in a very old
population. Small differences between the light fractions in the
oldest populations (which are poorly constrained at the $\lesssim
10$\% level) for the VLT and STIS fits generate large differences in
$\log_{10} \langle \tau \rangle$ ($1.97$ dex for NGC 1493 and
$-0.82$ dex for NGC 2139). By contrast, a statistic that is less
sensitive to small amounts of light in the oldest population show
much better agreement between VLT and STIS. For example, the
difference in $\langle \log_{10} \tau \rangle$ (the
luminosity-weighted average of $\log_{10} \tau$ for the composite
fits) between the two samples is $0.31 \pm 0.18$ dex. This has about
three times less scatter than a comparison of $\log_{10} \langle
\tau \rangle$.

The conclusions that we can draw from these comparisons are somewhat
limited. Either way, it does not appear that the data provide
evidence for the presence of strong population gradients between
spatial resolutions of $0.2"$ (which mostly sample the NC) and
$1.0"$ (which samples both the NC and the galaxy disk). Therefore,
we believe that our VLT results provide a reliable characterization
of the NC properties in our sample galaxies, despite the
contamination from disk light.

%%%%%%%%%%%%%%%%%%%%%%%%%%%%%%%%%%%%%%%%%%%%%%%%%%%%%%%%%%%%%
\section{Discussion} 
\label{s:disc}

In summary, our extensive population analysis has shown that there 
is strong evidence that nuclear clusters have multiple-aged stellar
populations. The average metallicity is somewhat sub-solar, but
is higher that those of disk globulars in the Milky Way.
The derived properties of the nine NCs in our 
sample are summarized in Table \ref{t:prop}. The mean age 
$\log\, \langle \tau \rangle$ and derived metallicity $Z$ are taken 
from Section \ref{s:mixpop}. 
We argued in Section \ref{s:age_methods} that these should be the
determinations least influenced by systematic biases. The 
age of the most recent burst of star formation log($\tau_{lb})$ is taken from 
Section \ref{s:lastburst}. We have not listed formal errors on these 
quantities, as the uncertainties are dominated by systematic and 
modelling uncertainties, as extensively discussed in Section 
\ref{s:age}. We give order of magnitude estimates in the footnotes to the 
Table. The masses log(M) in Table \ref{t:prop} are dynamical mass 
estimates, taken from Paper I. There masses were derived from 
a direct measurement of the velocity dispersion and careful 
dynamical modelling, which is much less prone to systematic 
problems than an analysis based on population synthesis models.

%%%%%%%%%%%%%%%%%%%%%%%%%%%%%%%%%%%%%%%%%%%%%%%%%%%%%%%%%%%%%
\subsection{The Duty Cycle of Star Formation in NCs}
\label{s:dutycycle}

The age of the last star formation burst in the NCs can be used 
to derive the duty cycle of star formation in the clusters, i.e. 
the fraction of time during which the clusters are 
actively forming stars. For each cluster we know that no sizable 
star formation occurred in the time since the last burst. So, each 
individual cluster was on a fraction $f^k_{on} = T_{\mbox{on}}/\tau^k_{lb}$ 
of the time, where $T_{\mbox{on}}$ is the typical duration of a 
star formation episode. Here $\tau^k_{lb}$ is the time since 
the last burst in cluster $k$ as taken from Table \ref{t:prop}. 
All $n$ clusters together were on a fraction 

\begin{equation}
 f_{on} = \frac{n T_{\mbox{on}}}{\sum_{k=1}^n \tau_{lb}^k}
\end{equation}
of the time, which is the current duty cycle. Weidner et al. (2004) 
argue that the typical formation timescale $T_{\mbox{on}}$ 
of massive star 
clusters is between $10^6$ and $10^7$ years. If we assume 
the typical duration of one star formation event is $10^6$ years, 
the duty cycle as derived from the nine clusters in our sample is 
3 \%. Assuming the typical duration of a burst is $10^7$ years 
yields a duty cycle of 30 \%.

A useful comparison number that we will elaborate on in a future paper 
on emission line properties of the NCs is the number 
of spectra that show prominent emission lines that could be indicative 
of ongoing star formation. Remember, however, that the spectra sample 
more than just the NC light in most of the cases and 
we have at present no means to be sure whether the emission really 
comes from the cluster or from the surrounding disk. It also remains 
possible that low-luminosity AGN contribute to the observed emission 
line flux. Out of the nine 
spectra, three show no emission, five show prominent emission and one is, 
though clearly detected, not very strong. If one were to derive a 
duty cycle from these numbers, it thus turns out to be around 60 \%. 
Though we may thus have overestimated the age of the last star 
formation burst, this only strengthens our conclusion that the 
centers of late-type spirals are regularly forming new stars. 

It should be noted here that the relatively low average 
metallicities we derive 
are consistent with the recurrent star formation scenario. In 
fact the metallicity of a recurrently replenished closed box 
(i.e. no outflow of metals from feedback winds, but inflow of 
fresh gas) 
would be constant with time and depends only on the star formation 
efficiency and yield in the NC and on the metallicity of the infalling gas.

%%%%%%%%%%%%%%%%%%%%%%%%%%%%%%%%%%%%%%%%%%%%%%%%%%%%%%%%%%%%%
\subsection{The Past Star Formation Rates of NCs}
\label{s:SFR}

We also derive the star formation rate $<\mbox{SFR}>$ in the 
mean over eight of the nine clusters in our sample. 
NGC~7418, which has by far the largest mass of all NCs and also 
a sizable percentage of young stars, would contribute the bulk 
of the recent star formation to a simple average although it 
may not be typical for NCs as a class. We thus exclude it 
from the following averages. We follow 
two different approaches. First we simply assume that 
clusters build their mass $M_{NC}$ (as given in Table 
\ref{t:prop}) continually over a Hubble time $t_H = 13$ Gyr:

\begin{equation}
 <\mbox{SFR}>_1 = \sum_{k=1}^8 \frac{M_{NC}^k}{8 \times t_H} \approx 2\times10^{-4} M_{\odot}/\mbox{yr}.
\end{equation}

Second we can check if the SFR is constant over $t_H$ by comparing it to 
the SFR over the most recent $10^8$ years. We take the best fit 
mass weights $m^k_l$ of the composite population fits 
at face value. Here $k$ is an index running over the eight 
clusters and $l$ is an index running over the 14 SSP templates.
We consider two broad age bins: the recently 
formed mass with age between $10^6$ and $10^8$ years and the old 
or ``underlying'' population with age between $3\times10^8$ 
years and 20 Gyr. The mean star formation rate $<\mbox{SFR}>_2$ in 
the young bin is then

\begin{equation}
 <\mbox{SFR}>_2 = \sum_{k=1}^8 \left( \frac{\sum_{l=1}^7 m^k_l  M_{NC}^k}{8 \times 10^8 \mbox{yr}} \right) = 2.4 \times 10^{-3} M_{\odot}/\mbox{yr} , 
\end{equation}
where $l$ runs from the template with age $10^6$ years and index 
$l$ = 1 to the template with age $10^8$ years and index $l$ = 7.

We found in Section \ref{s:lastburst} that the star formation 
bursts are typically separated by $2 \times 34$ Myr (the factor two 
arises because the probability to observe one object at any time 
between two bursts is constant with time). Thus in the mean 
over all objects the most likely point in time at which we 
observe is when half of the time between two bursts has elapsed 
since the last burst. The mass of 
stars that is formed per burst $M_{fpb}$ is then 

\begin{equation}
 M_{fpb} =  <\mbox{SFR}>_2 \times 2 \times 34 \mbox{Myr} = 1.6 \times 10^{5} M_{\odot}.
\end{equation}
This number can be compared with the mean mass per burst derived 
in Section \ref{s:lastburst}. Leaving out NGC~7418, one finds 
M = $1.9 \times 10^{5} M_{\odot}$ for this comparison number. 
Each cluster has thus experienced of the order of 25 bursts in 
its lifetime. 

As discussed in Section \ref{s:age_methods}, the mass fractions of 
the populations in each cluster are subject to considerable scatter. 
Therefore the star formation rates we derive could be uncertain by up 
to a factor of 2. 
However $<\mbox{SFR}>_2$ is bigger than $<\mbox{SFR}>_1$ 
by one order of magnitude. At the 
current $<\mbox{SFR}>_2$ (leaving out NGC~7418), NCs need only be 
$\approx$ 2 Gyr old to build up their typical mass of $5\times 10^6$ 
M$_{\odot}$. Assuming that current star formation rates 
are typical for the whole existence of the 
clusters, would lead to the conclusion that NCs 
do not need to form in the very young universe. We already 
pointed out in Section \ref{s:lastburst} that the NC 
of NGC~2139 possibly contains no population older than 100 Myr. 
NGC~2139 is also the lightest of the clusters at a mass of 
only $8 \times 10^5$ M$_{\odot}$ (one could speculate that this 
could be the typical mass at formation time for a NC). 
Note that the mass we derive as the mass formed per burst 
is close to the mass we derive for this NC. 
So it is possible that it formed just recently, i.e. around 
$10^8$ years ago. 

On the other hand it should be remembered that the mass weights 
of the composite fits in Table \ref{t:mix_weights} show that 
out of nine clusters, six formed more than 50\% of their mass more 
than $10^{9.4}$ years ago. There is also no basis for the assumption 
that the star formation rate remains constant over time. Indeed, 
more massive clusters could attract more gas, therefore increasing 
their star formation rate over their lifetime. Finally, we are biased 
observationally to miss the very faintest, i.e. possibly older, NCs. 
Therefore we are possibly biased to higher SFRs.

%%%%%%%%%%%%%%%%%%%%%%%%%%%%%%%%%%%%%%%%%%%%%%%%%%%%%%%%%%%%%
\subsection{Clues on the Formation Process?}
\label{s:form}

As stated in the introduction, there is no obvious, compelling 
reason why the centers of late-type spirals should be conducive 
to the genesis of such massive, compact clusters. 
In standard star formation processes, the upper limit 
of the cluster mass distribution depends on the star formation rate 
in the sense that higher star formation rates lead to the formation 
of a larger number of clusters, which then sample further into the 
high-mass tail of the cluster mass distribution function (Larsen 2002). 
Star formation rates in late-type disks are in no way comparable 
to those of starburst galaxies like M82, where the formation of 
massive clusters has been observed (e.g. McCrady et al. 2003). 
Cluster formation in starburst galaxies is also expected to 
be a spatially random process, in clear contradiction to the 
observation that there is only one NC, which is located in the 
very nucleus of its host galaxy. 
It is thus highly unlikely that NCs form from such a standard 
cluster formation mode. Furthermore, though repetitive star 
formation in galactic nuclei has been predicted by e.g. 
Kr\"ugel \& Tutukov (1993), these results are not 
directly applicable to the present case as their scenario 
includes the steep potential well of a bulge and spatial 
scales of up to 1 kpc, far in excess of the typical size 
of the NC. 

A few authors have provided interesting first steps towards understanding 
the formation of NCs. Under the assumption 
of a singular gravitational potential, Milosavljevic 
(2004) has shown that the disks of late-type spirals can 
transport gas inwards at a rate of roughly $1 \times 10^{-2}$ 
M$_{\odot}$/yr, which is sufficient to sustain the star formation rates we 
derive in the present paper. It remains however unclear, why the star 
formation should concentrate in a spatial region with a size of only 5 pc. 
Bekki et al. (2004) and Fellhauer et al. (2002) consider the case in which 
massive star clusters are formed from the merger of several recently 
formed ``normal'' star clusters under the influence of their own gravity. 
Typical initial conditions are adapted to regions of very high star 
formation activity, as this is the environment where it is most likely 
to find a sufficient number of star clusters (around 20) in a 
sufficiently small volume (a size of roughly 100 pc). 
The process can happen on a timescale of less than $10^8$ years if the 
initial conditions are right. If the process was to go much slower 
(more than a Gyr), the still distinct clusters in the process of 
merging should be visible in the HST images, which is not the case 
(compare B02). The effective radii they obtain from these simulations 
are between 10 and 50 pc. While this scenario could explain high 
masses, it thus generally fails to reproduce the small sizes and mixed 
populations of the observed NCs.

The origin of the nuclei of dE galaxies, which are possibly 
related to NCs, is generally attributed to two 
possible mechanisms, i.e. 
(a) the decay of the orbit of a pre-existing globular cluster toward 
the bottom of the galactic potential well, driven by dynamical friction, or 
(b) the in situ formation of a giant cluster from gas fallen to 
the center of the galaxy (see e.g. Durrell 1997). Dynamical friction 
timescales (larger than several Gyrs, see Milosavljevic 2004) 
are too long to account for the generally young ages 
of the NCs in late-type spirals. As 
a conjecture as for how NCs form, the following 
scenario would however fit all observed properties: Take a randomly 
forming star cluster in the very vicinity of the galaxy center 
(i.e. center of rotation and potential). Once formed, this cluster's 
potential will itself be the minimum of the total galaxy's potential 
(or it will tend to drift their because of dynamical friction). 
Thus the gas present in the disks of extremely late-type 
spirals will be accreted on the cluster. The gas can reach the center 
in a quiet, non-star forming mode because the surface density 
of the disk is so low. Thus the gas is only induced to form stars 
when falling into the proto-NC. It thus contributes 
to the high space density in the cluster. As a large fraction of the 
infalling gas will thus reach the NC, it can grow ever 
bigger. This scenario thus could explain the high masses, 
small radii and recurring star formation events in the clusters. 
It also describes in a natural way, why 25\% of the late-type 
spirals lack a NC: simply because they have up to now lacked a 
suitable seed cluster. 
Whether this scenario can quantitatively explain the observed 
properties of NCs remains to be tested with detailed numerical 
simulations. It remains unclear in particular, how well the 
coincidence between NC and galaxy center can be accounted for.

%%%%%%%%%%%%%%%%%%%%%%%%%%%%%%%%%%%%%%%%%%%%%%%%%%%%%%%%%%%%%
\section{Conclusions}
\label{s:concl}

We have presented results from a spectroscopic survey
of the nuclear regions of late-type spiral galaxies, undertaken
with VLT/UVES. We aimed to study the stellar population properties 
of the single, most luminous star cluster in the photometric 
center of the galaxy, which we call a nuclear star cluster. 
We used population synthesis models 
to derive mean luminosity weighted ages within the aperture of 
the extracted spectra. We found no evidence for strong radial age 
gradients between spatial resolutions of $0\farcs2$ (which mostly 
sample the nuclear cluster, from HST/STIS data) and $1''$ 
(which samples both the nuclear cluster and the galaxy disk, from 
VLT/UVES data), implying that we can directly 
measure the stellar populations of the nuclear clusters.
Fits using simple SSP models yield ages 
of the dominating population in the spectra that range between  
$1\times 10^7$, and  $3\times 10^9$ years, and which are robust to 
the assumed metallicity or internal extinction. They 
are also consistent between two different approaches, 
namely either using Lick/IDS indices or directly fitting the 
spectrum with templates from the model over a spectral 
range of 1100{\AA}.

We also fitted for the age composition of the clusters. 
These composite fits show that the luminosity weighted 
mean ages of nuclear clusters actually range from $4.1\times 10^7$ to 
$1.1\times 10^{10}$ years, with uncertainties of 0.3 dex.
These ages are older than those suggested by SSP models. 
The average metallicity of nuclear clusters in late-type spirals 
is slightly sub-solar ($\langle Z \rangle = 0.015$) but shows 
significant scatter. This is lower than for the nuclear regions 
in earlier-type galaxies (see e.g., Sarzi et al. 2005 and 
Rossa et al. 2006). The population synthesis model, 
however, does not at present cover varying abundance ratios, 
which limits the accuracy of this metallicity determination. 
While most of the clusters 
have moderate extinctions of 0.1 to 0.3 mags in the $I$-band, 
two out of nine clusters are extincted by 0.55 and 0.85 mag, 
respectively. The stellar mass-to-light ratios 
were well constrained from the composite-age fits and agree with 
the dynamical mass estimates. However a residual 
scatter remains due to the uncertain contributions of the oldest 
populations to the total light. 
We were also able to derive the age of the last star 
formation episode, which was in the mean over our nine 
sample clusters 34 Myr ago. All of the clusters 
experienced star formation in the last 100 Myr. 
For the clusters in our sample, 
an estimate for the mean star formation rate 
over the last 100 Myr is $<\mbox{SFR}> =2 \times 10^{-3}M_{\odot}/{yr}$. 

We have presented a number of independent arguments that point 
towards repetitive star formation in the nuclear clusters:

\begin{itemize}
\item The mean age of light dominating the clusters is $10^8$ years, 
which is much smaller 
than the age of the universe. Excluding the possibility that we live in 
special times, i.e. that all clusters happened to form recently, this 
can be explained convincingly in a repetitive scenario, as younger 
stellar populations have intrinsically lower mass-to-light ratio 
than older populations. Thus they mask the main mass of the cluster 
which resides in a significantly older, ``underlying'' population.
It has to be 
kept in mind however that selection effects might bias us 
towards younger clusters as we only sample the brighter 2/3 
of the luminosity range of the nuclear clusters.

\item The measured spectral indices are consistent with a continuous 
star formation history and inconsistent with a single burst in 
most cases. 

\item Single-age fits to the spectra are significantly worse in a 
$\chi^2$ sense than the multi-component fits in all cases.

\item The cluster mass-to-light ratios as derived from the 
single-age fits are not consistent with those derived from 
the dynamical modelling of Paper I, while those derived from 
the composite fits are. 

\end{itemize}

The population fits and star formation rates show that nuclear clusters form 
stars onto the present day. A caveat remains, namely that some of the
mixed populations that we see might be due to disk star
contamination in the spectra. However we found no evidence 
for strong radial age gradients, hence no evidence for strong 
disk contamination.

Taking these results together with those from Paper I we 
have thus shown that nuclear star clusters are massive and 
dense star clusters that form stars recurrently until the 
present day. We also refer to the recent paper of Rossa 
et al. (2006) for a study on a larger sample of NCs, which 
finds results that are entirely consistent with ours. A detailed 
comparison between our two samples is also given in that paper. 
While nuclear clusters are dynamically and structurally similar 
to the most massive globular clusters and super-star-clusters, 
these results show that their star formation histories and 
populations are unique in the whole star cluster domain.
It is almost inevitable to associate these unique properties 
with the location of the cluster in its host galaxy. It 
remains a challenging question to elucidate exactly how 
very late-type spirals manage to create their extreme 
property nuclei. In turn, our results make perhaps even more 
puzzling why other, structurally similar galaxies, seem to have 
no significant nuclear clusters at all (B04).

%%%%%%%%%%%%%%%%%%%%%%%%%%%%%%%%%%%%%%%%%%%%%%%%%%%%%%%%%%%%%
\section*{Acknowledgments}       

Support for the HST data, proposal \#9070, was provided by NASA 
through a grant 
from the Space Telescope Science Institute, which is operated by the 
Association of Universities for Research in Astronomy, Inc., under NASA 
contract NAS 5-26555. CJW and HWR wish to thank the observing staff 
at the VLT for their excellent support. 
We thank the anonymous referee for suggestions that helped
improve the presentation of the paper.

%%%%%%%%%%%%%%%%%%%%%%%%%%%%%%%%%%%%%%%%%%%%%%%%%%%%%%%
%%%%% References %%%%%

\clearpage

\clearpage

\clearpage

\clearpage

\clearpage

\clearpage

\clearpage

\clearpage

\clearpage

\clearpage

\clearpage

\clearpage

\clearpage

\begin{deluxetable}{ccccccc}
\tablecaption{Sample properties}
\tablehead{
{Galaxy}  & Hubble & {Distance}& {M$_{I}$} &  Aperture  & r$_e$       & {NCL} \\
          &  type  & {[Mpc]}   &           & [$\arcsec$]& [$\arcsec$] & {[\%]}\\
    (1)   &  (2)   &     (3)   &   (4)     &    (5)     &  (6)        & (7)   \\
}
\startdata
NGC~300   & SAd    & 2.2       &  $-11.43$ &  1.09      & 0.272 &  0 \\
NGC~428   & SABm   & 16.1      &  $-13.15$ &  0.93      & 0.043 & 30 \\
NGC~1042  & SABcd  & 18.2      &  $-12.95$ &  0.93      & 0.022 & 81 \\
NGC~1493  & SBcd   & 11.4      &  $-11.43$ &  1.24      & 0.047 & 33 \\
NGC~2139  & SABcd  & 23.6      &  $-12.65$ &  1.24      & 0.090 & 89 \\
NGC~3423  & SAcd   & 14.6      &  $-11.84$ &  0.93      & 0.059 & 76 \\
NGC~7418  & SABcd  & 18.4      &  $-16.23$ &  1.40      & 0.140 & 18 \\
NGC~7424  & SABcd  & 10.9      &  $-11.41$ &  1.55      & 0.140 & 77 \\
NGC~7793  & SAd    & 3.3       &  $-13.64$ &  1.55      & 0.484 &  0 \\
\enddata
\label{t:UVES}
\tablecomments{Cols. (1) and (2) Galaxy name and Hubble classification, 
as taken from NED. 
Col. (3) Distances were taken from B02, where they were calculated from 
the recession velocity (from LEDA, corrected for virgocentric infall) 
and assume H$_0 = \rm{70\,km\,s^{-1}\,Mpc^{-1}}$. Col. (4) Absolute magnitude 
of the NC as taken from B02. Col. (5) Width of the extraction aperture 
in spatial direction. Col. (6) and (7) Effective radius of the NC and 
contamination from 
galaxy light in percent, as derived in Paper I.}
\end{deluxetable}

\clearpage

\begin{deluxetable}{cccccccccc}
\tablecaption{Measured indices}
\rotate
\tablewidth{0pt}
\tabletypesize{\scriptsize}
\tablehead{
\colhead{NGC No.} &
\colhead{0300} &
\colhead{0428} &
\colhead{1042} &
\colhead{1493} &
\colhead{2139} &
\colhead{3423} &
\colhead{7418} &
\colhead{7424} &
\colhead{7793}
}
\startdata
D$_n$(4000)  & 1.534 $\pm$ 0.005  & 1.85 $\pm$ 0.03 &  1.408 $\pm$ 0.008  & 1.231 $\pm$ 0.002 & 1.042 $\pm$  0.005 &  1.632 $\pm$ 0.005  & 1.317 $\pm$  0.005 & 1.223 $\pm$ 0.005 & 1.317 $\pm$ 0.005\\
B4000        & 1.369 $\pm$ 0.0056 & 1.57 $\pm$ 0.03 &  1.27 $\pm$ 0.01    & 1.130 $\pm$ 0.002 & 1.011 $\pm$  0.006 &  1.483 $\pm$ 0.006 & 1.157 $\pm$  0.006 & 1.112 $\pm$ 0.007 & 1.182 $\pm$ 0.001\\
H$\delta_A$    & 4.4 $\pm$ 0.2      & 2.7 $\pm$ 0.8   &  3.6 $\pm$ 0.3      & 5.84 $\pm$ 0.09   & 5.3 $\pm$  0.3     &  1.9 $\pm$ 0.2  & 4.7 $\pm$  0.2 & 6.4 $\pm$ 0.3 & 7.29 $\pm$ 0.03\\
Ca4227       & 0.6 $\pm$ 0.1      & 0.1 $\pm$ 0.4   &  0.8 $\pm$ 0.2      & 0.20 $\pm$ 0.05   & 0.0 $\pm$  0.2     &  1.45 $\pm$ 0.09 & 0.3 $\pm$  0.1 & 0.4 $\pm$ 0.1 & 0.45 $\pm$ 0.02\\
G4300        & 2.5 $\pm$ 0.2      & 1.6 $\pm$ 0.7   &  2.2 $\pm$ 0.4      & 0.1 $\pm$ 0.1     & 0.1 $\pm$  0.4     &  3.7 $\pm$ 0.2 & 0.0 $\pm$  0.3 & -0.3 $\pm$ 0.3 & 0.23 $\pm$ 0.04\\
    H$\gamma$   & 1.1 $\pm$ 0.2 & 0.0 $\pm$ 0.7 &  0.0 $\pm$ 0.3 & 4.0 $\pm$ 0.1 & 4.8 $\pm$  0.3 & -2.2 $\pm$ 0.2 & 2.6 $\pm$  0.2 & 4.6 $\pm$ 0.3 & 4.88 $\pm$ 0.04\\
      Fe4383    & 2.0 $\pm$ 0.2 & 1.6 $\pm$ 0.9 &  3.2 $\pm$ 0.5 & 0.3 $\pm$ 0.1 & 1.0 $\pm$  0.5 &  4.5 $\pm$ 0.2  & 1.7 $\pm$  0.4 & 1.4 $\pm$ 0.4 & 1.00 $\pm$ 0.06\\
       Ca4455   & 1.0 $\pm$ 0.1 & 0.0 $\pm$ 0.4 &  1.2 $\pm$ 0.2  & 0.24 $\pm$ 0.07 & 1.4 $\pm$  0.3 &  1.8 $\pm$ 0.1  & 1.1 $\pm$  0.2 & 1.0 $\pm$ 0.2 & 0.52 $\pm$ 0.03\\
      Fe4531    & 2.3 $\pm$ 0.2 & 2.2 $\pm$ 0.7 &  2.5 $\pm$ 0.4  & 1.4 $\pm$ 0.1 & 0.5 $\pm$  0.5 &  2.5 $\pm$ 0.2 & 0.8 $\pm$  0.3 & 1.1 $\pm$ 0.4 & 1.46 $\pm$ 0.05\\
     Ca\textsc{II}8542   & 2.9 $\pm$ 0.1 & 1.9 $\pm$ 0.3 &  3.7 $\pm$ 0.1  & 2.9 $\pm$ 0.2 & 2.5 $\pm$  0.1 &  3.1 $\pm$ 0.2 & 3.1 $\pm$  0.1 & 2.4 $\pm$ 0.2 & 3.4 $\pm$ 0.1\\
     Ca\textsc{II}8662   & 2.4 $\pm$ 0.1 & 2.3 $\pm$ 0.3 &  3.6 $\pm$ 0.1  & 3.1 $\pm$ 0.2 & 2.8 $\pm$  0.1 &  2.2 $\pm$ 0.2 & 3.1 $\pm$  0.1 & 2.4 $\pm$ 0.2 & 3.34 $\pm$ 0.09 \\
\enddata
\label{t:measind}
\tablecomments{The following values are obtained for the Balmer indices if 
no interpolation of the emission lines is performed. \\
NGC~1042: H$\delta_A$ = 3.3 $\pm$ 0.3, H$\gamma$ = -0.8 $\pm$ 0.4  \\
NGC~7418: H$\delta_A$ = 4.4 $\pm$ 0.2, H$\gamma$ = 1.5 $\pm$  0.2\\
NGC~7424: H$\delta_A$ = 6.3 $\pm$ 0.3, H$\gamma$ = 4.0 $\pm$  0.3\\
}
\end{deluxetable}

\clearpage

\begin{deluxetable}{crcc}
\tablecaption{Ages and metallicities from absorption indices} 
\tablehead{
Galaxy     &   $\chi^2$  & log($\tau$)    & $Z$ \\
}
\startdata
     NGC~ 300    &  154  &  9.11$_{-0.00}^{+0.10}$ & 0.004 \\
     NGC~ 428    &  4.2  &  9.36$_{-0.26}^{+0.08}$ & 0.004 - 0.02 \\
     NGC~1042    &  225  &  8.00$_{-1.00}^{+1.00}$ & 0.004 - 0.05 \\
     NGC~1493    &  69   &  7.74$_{-0.05}^{+0.11}$ & 0.004 \\
     NGC~2139    &  6    &  7.30$_{-0.19}^{+0.06}$ & 0.05 \\
     NGC~3423    &  194  &  9.39$_{-0.24}^{+0.36}$ & 0.004 - 0.02 \\
     NGC~7418    &  75   &  7.70$_{-0.21}^{+0.16}$ & 0.004  \\
     NGC~7424    &  1.8  &  7.81$_{-0.05}^{+0.05}$ & 0.004 \\
     NGC~7793    &  1417 &  8.26$_{-0.15}^{+0.65}$ & 0.004 - 0.008 \\
\enddata
\label{t:ind_age}
\tablecomments{NGC~7418 has two minima that are almost equally good 
fits, the other minimum is at log($\tau$) = 6.86$_{-0.06}^{+0.05}$ 
and $Z$=0.008. As the coverage of 
the models in $Z$ is sparse, the uncertainties in metallicity 
are noted as two different possible values in some cases.}
\end{deluxetable}

\clearpage

\begin{deluxetable}{c|rcccc}
\tablecaption{Ages and metallicities from fitting SSPs} 
\tablehead{
Galaxy           &$\chi^2$ &  $Z$    & A$_I$ &  log($\tau$) & M/L$_I$ \\
}
\startdata
     NGC~300     &  4128  &  0.004   &  0.35  &   9.00  &   0.41 \\
     NGC~428     &   246  &  0.05    &  0.25  &   9.00  &   0.41 \\
     NGC~1042    &  1107  &  0.004   &  1.35  &   7.48  &   0.07 \\
     NGC~1493    &  4604  &  0.008   &  0.60  &   7.48  &   0.07 \\
     NGC~2139    &   741  &  0.008   &  0.00  &   7.00  &   0.02 \\
     NGC~3423    &  6519  &  0.004   &  0.00  &   9.48  &   0.87 \\
     NGC~7418    &  1509  &  0.004   &  0.55  &   7.78  &   0.10 \\
     NGC~7424    &   594  &  0.004   &  0.20  &   7.78  &   0.10 \\
     NGC~7793    & 34768  &  0.008   &  0.30  &   8.00  &   0.15 \\
\enddata
\label{t:SSP_age}
\end{deluxetable}

\clearpage

\begin{deluxetable}{c|rccccc}
\tablecaption{Ages and metallicities from composite age fits} 
\tablehead{
Galaxy     & $\chi^2$ &  $Z$  & A$_I$ & $\log \langle \tau_I \rangle$ & M/L$_I$ \\
}
\startdata
NGC~300     & 1248  &   0.004    &    0.15  &    9.49   &   0.64\\
NGC~428     &  189  &   0.02     &    0.85  &    9.27   &   0.60\\
NGC~1042    &  320  &   0.02     &    0.30  &   10.05   &   2.06\\
NGC~1493    & 3996  &   0.008    &    0.30  &    9.76   &   0.88\\
NGC~2139    &  739  &   0.05     &    0.00  &    7.61   &   0.08\\
NGC~3423    & 3956  &   0.008    &    0.15  &    9.75   &   1.12\\
NGC~7418    &  737  &   0.008    &    0.55  &    9.05   &   0.30\\
NGC~7424    &  365  &   0.008    &    0.15  &    9.11   &   0.31\\
NGC~7793    &27281  &   0.008    &    0.00  &    9.29   &   0.47\\
\enddata
\label{t:mix_age}
\tablecomments{There are N$_{\mbox{DOF}}$ = N$_{\mbox{pix}}-$N$_{\mbox{param}} \approx 900$ degrees of freedom.}
\end{deluxetable}

\begin{deluxetable}{c|cccccccccccccc}
\tablecaption{Linear mass weights of contributing SSPs in the composite fit}
\tablehead{& & & & & & & NGC~ & & &\\
$\log$(age) & M/L$_I$  & 300 & 428 & 1042 & 1493 & 2139 & 3423 & 7418 & 7424 & 7793 \\
}
\startdata
6.00 & 0.05 & 0.000 & 0.000 & 0.000 & 0.000 & 0.000 & 0.001 & 0.009 & 0.000 & 0.000  \\
6.48 & 0.02 & 0.000 & 0.000 & 0.000 & 0.000 & 0.079 & 0.000 & 0.000 & 0.000 & 0.000  \\
6.78 & 0.02 & 0.000 & 0.001 & 0.001 & 0.000 & 0.000 & 0.000 & 0.004 & 0.000 & 0.000  \\
7.00 & 0.02 & 0.000 & 0.000 & 0.000 & 0.004 & 0.000 & 0.000 & 0.000 & 0.019 & 0.000  \\
7.48 & 0.07 & 0.000 & 0.000 & 0.000 & 0.031 & 0.000 & 0.002 & 0.087 & 0.059 & 0.012  \\
7.78 & 0.10 & 0.000 & 0.000 & 0.000 & 0.000 & 0.921 & 0.000 & 0.000 & 0.000 & 0.000  \\
8.00 & 0.15 & 0.027 & 0.000 & 0.004 & 0.000 & 0.000 & 0.000 & 0.000 & 0.000 & 0.101  \\
8.48 & 0.23 & 0.000 & 0.000 & 0.000 & 0.014 & 0.000 & 0.000 & 0.056 & 0.291 & 0.079  \\
8.78 & 0.34 & 0.000 & 0.000 & 0.000 & 0.043 & 0.000 & 0.000 & 0.117 & 0.000 & 0.045  \\
9.00 & 0.41 & 0.013 & 0.348 & 0.052 & 0.000 & 0.000 & 0.000 & 0.154 & 0.000 & 0.000  \\
9.48 & 0.87 & 0.637 & 0.651 & 0.019 & 0.000 & 0.000 & 0.000 & 0.000 & 0.000 & 0.000  \\
9.78 & 1.47 & 0.323 & 0.000 & 0.000 & 0.000 & 0.000 & 0.997 & 0.574 & 0.000 & 0.762  \\
10.00 & 2.08 & 0.000 & 0.000 & 0.000 & 0.000 & 0.000 & 0.000 & 0.000 & 0.631 & 0.000  \\
10.30 & 3.53 & 0.000 & 0.000 & 0.924 & 0.908 & 0.000 & 0.000 & 0.000 & 0.000 & 0.000  \\
\enddata
\label{t:mix_weights}
\tablecomments{The mass weights are the fraction of the total 
stellar mass at the present time.}
\end{deluxetable}

\clearpage

\begin{deluxetable}{l|lrrr|lrrr}
\tablecaption{Population Analysis Comparison VLT vs.~STIS}
\tablehead{
Galaxy & $Z$ & $A_I$ & $\log_{10}\tau_{\rm VLT}$ & 
             $\log_{10}\tau_{\rm STIS}$ & 
         $Z$ & $A_I$ & $\log_{10}\langle\tau\rangle_{\rm VLT}$ & 
             $\log_{10}\langle\tau\rangle_{\rm STIS}$ \\
(1) & (2) & (3) & (4) & (5) & (6) & (7) & (8) & (9) \\
}
\startdata
NGC\,300  & 0.004 & 0.35& 9.00 & 8.48 & 0.004 & 0.15& 9.49 & 9.50 \\ 
NGC\,428  & 0.05  & 0.25& 9.00 & 8.78 & 0.02  & 0.85& 9.27 & 8.70 \\
NGC\,1493 & 0.008 & 0.6 & 7.48 & 7.00 & 0.008 & 0.30& 9.76 & 7.79 \\
NGC\,2139 & 0.008 & 0.0 & 7.00 & 7.00 & 0.05  & 0.0 & 7.61 & 8.43 \\
\enddata
\label{t:STIS}
\tablecomments{Col.~(1) lists the galaxy name. Cols.~(2)--(4) list the results
from SSP fits to the VLT data for the metallicity, $I$-band extinction,
and age, respectively (from Table \ref{t:SSP_age}). Col.~(5)
list the results from SSP fits to the STIS spectra, with metallicity
and $I$-band extinction fixed to the values inferred from the VLT
data. Cols.~(6)--(8) list the results from composite-age fits to the
VLT data for the metallicity, $I$-band extinction, and
luminosity-weighted mean age in the $I$-band,
 respectively (from Table \ref{t:mix_age}). 
Col.~(9) list the results from composite-age fits to the STIS
data, with metallicity and $I$-band extinction fixed to the values
inferred from the VLT data.}
\end{deluxetable}

\clearpage

\begin{deluxetable}{cccccc}
\tablecaption{Summary of properties} 
\tablehead{
Galaxy     & $\log\, \langle \tau \rangle$ & $Z$ & log($\tau_{lb})$ & {log(M) [M$_{\sun}$]} \\
}
\startdata
NGC~0300  & 9.49 & 0.004 & 8.00 & $6.02 \pm 0.24 $\\
NGC~0428  & 9.27 & 0.02  & 7.48 & $6.51 \pm 0.14 $\\
NGC~1042  & 10.05& 0.02  & 7.00 & $6.51 \pm 0.21 $\\
NGC~1493  & 9.76 & 0.008 & 7.00 & $6.38 \pm 0.14 $\\
NGC~2139  & 7.61 & 0.05  & 6.48 & $5.92 \pm 0.20 $\\
NGC~3423  & 9.75 & 0.008 & 7.00 & $6.53 \pm 0.14 $\\
NGC~7418  & 9.05 & 0.008 & 7.00 & $7.78 \pm 0.19 $\\
NGC~7424  & 9.11 & 0.008 & 7.48 & $6.09 \pm 0.14 $\\
NGC~7793  & 9.29 & 0.008 & 8.00 & $6.89 \pm 0.14 $\\
\enddata
\label{t:prop}
\tablecomments{Typical uncertainties on the mean ages $\log\, \langle 
\tau \rangle$ are $\pm$ 0.3 dex. Typical uncertainties on the 
metallicity $Z$ and the age of the youngest stellar population 
log($\tau_{lb})$ are half the step-size 
with which we sample the metallicity and age range, i.e. 
$\Delta Z \approx 0.2$ dex and $\Delta$ log($\tau_{lb}$) $\approx$ 0.15 dex 
respectively.}
\end{deluxetable}

\end{document}